\documentclass[11pt]{article}

    \usepackage[final]{acl}

    \usepackage{times}
    \usepackage{latexsym}

    \usepackage[T1]{fontenc}
    \usepackage[utf8]{inputenc}

    \usepackage{microtype}
    \usepackage{inconsolata}

    \usepackage{graphicx}

    \usepackage{tabularx}
    \usepackage{multirow}
    \usepackage{booktabs}
    \usepackage{amsmath}
    \usepackage{amssymb}
    \usepackage{array}
    \usepackage{subcaption}
    \usepackage{enumitem}
    \usepackage[table,dvipsnames]{xcolor}
    \usepackage[most]{tcolorbox}
    \usepackage{placeins}
    \usepackage{pifont}
    \usepackage{algorithm}
    \usepackage{algorithmic}
    \usepackage{listings}
    \usepackage{newfloat}
    \usepackage{caption}
    \usepackage{xurl}
    \definecolor{highlightgray}{gray}{0.85}
    \definecolor{tracegray}{RGB}{100,100,100}
    \definecolor{rationaleblue}{RGB}{0,102,204}
    \definecolor{errorred}{RGB}{220,53,69}
    \definecolor{correctgreen}{RGB}{40,167,69}
    \definecolor{initcolor}{RGB}{255,250,205}
    \definecolor{processcolor}{RGB}{230,240,255}
    \definecolor{finalcolor}{RGB}{230,255,230}

    \newcommand{\cmark}{\ding{51}}
    \newcommand{\xmark}{\ding{55}}
    \newcommand{\gcmark}{\textcolor{ForestGreen}{\cmark}}

    \lstdefinestyle{pythoncode}{
        language=Python,
        basicstyle=\ttfamily\small,
        keywordstyle=\color{blue}\bfseries,
        stringstyle=\color{red!70!black},
        commentstyle=\color{green!60!black}\itshape,
        numberstyle=\tiny\color{gray},
        showstringspaces=false,
        breaklines=true,
        frame=none,
        captionpos=b,
        tabsize=4,
        morekeywords={self,True,False,None},
    }

    \lstdefinestyle{tracestyle}{
        basicstyle=\ttfamily\scriptsize,
        backgroundcolor=\color{gray!10},
        frame=none,
        showstringspaces=false,
        breaklines=true,
    }

    \floatstyle{ruled}
    \newfloat{listing}{tb}{lst}
    \floatname{listing}{Listing}

    \newcolumntype{C}[1]{>{\centering\arraybackslash}p{#1}}

    \title{Think Like You Execute: Verifiable Chain of Thought from Program Traces}

    \author{
    Shailja Thakur$^1$, Vaibhav Saxena$^1$, Rohan Kulkarni$^1$, \\ 
    \textbf{Shivdeep Singh}$^1$, \textbf{Parameswaran Selvam}$^1$, \textbf{Hiroshi Kanayama}$^2$, \textbf{Hima Patel}$^1$ \\
    $^1$IBM Research, India $^2$IBM Research, Japan \\
    }

\begin{document}
    \maketitle

    \begin{abstract}
    Getting language models to reason correctly about code requires training on data where each reasoning step can be checked. Current synthetic Chain-of-Thought (CoT) training data often consists of plausible-sounding explanations generated by teacher models, and not verifiable accounts of actual program behavior. Models trained on such data learn logically flawed reasoning patterns despite syntactic correctness. To address this, we build a pipeline that generates execution-trace-verified CoT rationales by instrumenting code to capture traces, narrating them into natural language, and cross-checking each narration against the original trace. We systematically create 54,000 verified, bi-directional rationales that teach models to reason both forward (input$\rightarrow$output) and backward (output$\rightarrow$input). Models fine-tuned on our verified data achieve substantial improvements, with a peak gain of +26.6 on LiveCodeBench-Exec, +22.2 on CruxEval, and +19.5 on HumanEval across our fine-tuned models, demonstrating that verification quality directly determines both reasoning and code generation capabilities. Complete synthesis pipeline is avilable as open-source:\url{https://github.com/IBM/verified-code-cot/}
    \end{abstract}

    \section{Introduction}

    Recent advances in LLMs have enabled code assistants to generate, explain, and debug code~\cite{chen2021evaluating,austin2021program}, but they struggle when tasks require reasoning about what a program actually does~\cite{ding2024traced}---tracking how variables change during debugging, following execution paths to comprehend logic, or confirming that a refactored program preserves its original behavior. When asked to explain code execution, these models often produce plausible but incorrect accounts~\cite{turpin2023languagemodelsdontsay}.

    To improve reasoning capabilities, researchers have turned to Chain-of-Thought (CoT) training~\cite{wei2022chain}. However, current CoT datasets for code are generated without grounding in actual program execution, producing rationales that sound correct but are untethered to runtime behavior. Figure~\ref{fig:motivation} illustrates how LLMs hallucinate incorrect variable states and control flow despite having access to the code---for instance, comparing ``\texttt{arr[2]} with \texttt{arr[1]}'' instead of ``\texttt{arr[3]}'', and claiming peak direction as ``\texttt{right}'' instead of ``\texttt{left}''.

    \begin{figure*}[t!]
\centering\tiny\setlength{\tabcolsep}{1pt}\renewcommand{\arraystretch}{0.7}
\colorlet{phaseinit}{yellow!25}\colorlet{phaseloop}{blue!15}\colorlet{phasereturn}{green!20}\colorlet{llmbg}{red!12}\colorlet{halluctxt}{red!80!black}\colorlet{verifytxt}{green!50!black}\colorlet{tracebg}{black!8}
\begin{tabular}{@{}p{0.17\textwidth}|p{0.20\textwidth}|p{0.20\textwidth}|p{0.33\textwidth}@{}}
\toprule
\textbf{Code} & \textbf{LLM-Generated CoT} & \textbf{Execution Trace} & \textbf{Trace-Grounded \& Verified CoT (Ours)} \\
\midrule
{\ttfamily\tiny\textcolor{blue}{\textbf{def}} find\_peak(arr):\newline\hspace*{1mm}left,right=0,len(arr)-1\newline\hspace*{1mm}\textcolor{blue}{\textbf{while}} left<right:\newline\hspace*{2mm}mid=(left+right)//2\newline\hspace*{2mm}\textcolor{blue}{\textbf{if}} arr[mid]<arr[mid+1]:\newline\hspace*{3mm}left=mid+1\newline\hspace*{2mm}\textcolor{blue}{\textbf{else}}: right=mid\newline\hspace*{1mm}\textcolor{blue}{\textbf{return}} left\newline\textcolor{gray}{\# find\_peak([1,3,5,4,2])}}
&{\tiny\raggedright\cellcolor{llmbg} The function searches for a peak. With left=0, right=4, compute mid and check. \textcolor{halluctxt}{\textbf{At mid=2, compare arr[2] with arr[1]}} to determine direction. Since arr[2]>arr[1], peak to \textcolor{halluctxt}{\textbf{right}}, adjust until convergence.\newline\textcolor{halluctxt}{\ding{55} Hallucinates}}
&{\tiny\raggedright\cellcolor{tracebg}\sffamily\colorbox{phaseinit}{Init: left=0, right=4}\newline\colorbox{phaseloop}{Iter1: mid=2, arr[3]=4$\to$right=2}\newline\colorbox{phaseloop}{Iter2: mid=1, arr[2]=5$\to$left=2}\newline\colorbox{phasereturn}{Return: 2}}
&{\tiny\raggedright\colorbox{phaseinit}{\parbox{0.33\textwidth}{\raggedright Starting with array [1,3,5,4,2], we initialize left=0 and right=4.}}\\\vspace{-2pt}\colorbox{phaseloop}{\parbox{0.33\textwidth}{\raggedright First iteration finds midpoint at index 2 (value 5). Comparing with arr[3]=4 shows we're past the peak, so we set right=2.}}\\\vspace{-2pt}\colorbox{phaseloop}{\parbox{0.33\textwidth}{\raggedright Second iteration has mid=1 (value 3). Comparing with arr[2]=5 shows the peak is ahead, so left moves to 2.}}\\\vspace{-2pt}\colorbox{phasereturn}{\parbox{0.33\textwidth}{\raggedright Boundaries converge at index 2, which holds our peak value of 5.}}\\\vspace{1pt}\textcolor{verifytxt}{\ding{51} Verified step-by-step}}
\\
\bottomrule
\end{tabular}
\caption{Hallucinated vs trace-grounded CoT. LLM hallucinates (\texttt{arr[1]} vs \texttt{arr[3]}, \texttt{right} vs \texttt{left}). Our approach narrates execution traces with color correspondence (yellow=Init, blue=Loop, green=Return) proving verification.}
\label{fig:motivation}
\end{figure*}

    Prior work has tackled parts of this problem. Some translate execution traces into rationales~\cite{ding2024traced,armengol2025execution} but do not systematically verify the generated reasoning against the trace, while others check final outputs~\cite{li2025codeio,ding2024semcodertrainingcodelanguage,liu2025rstarcoderscalingcompetitivecode} but skip intermediate steps. Most also handle only forward reasoning (predicting outputs from inputs) and overlook backward reasoning (inferring inputs from outputs)~\cite{li2025codeio,jiang2024forward}. We close these gaps with an execution-grounded CoT synthesis pipeline that translates traces into natural language, cross-checks each rationale against the trace for (1) variable values, (2) state transitions, and (3) control flow, and generates bi-directional rationales for both forward and backward reasoning. 



    \begin{figure*}[t!]
        \centering
        \includegraphics[width=0.9\textwidth]{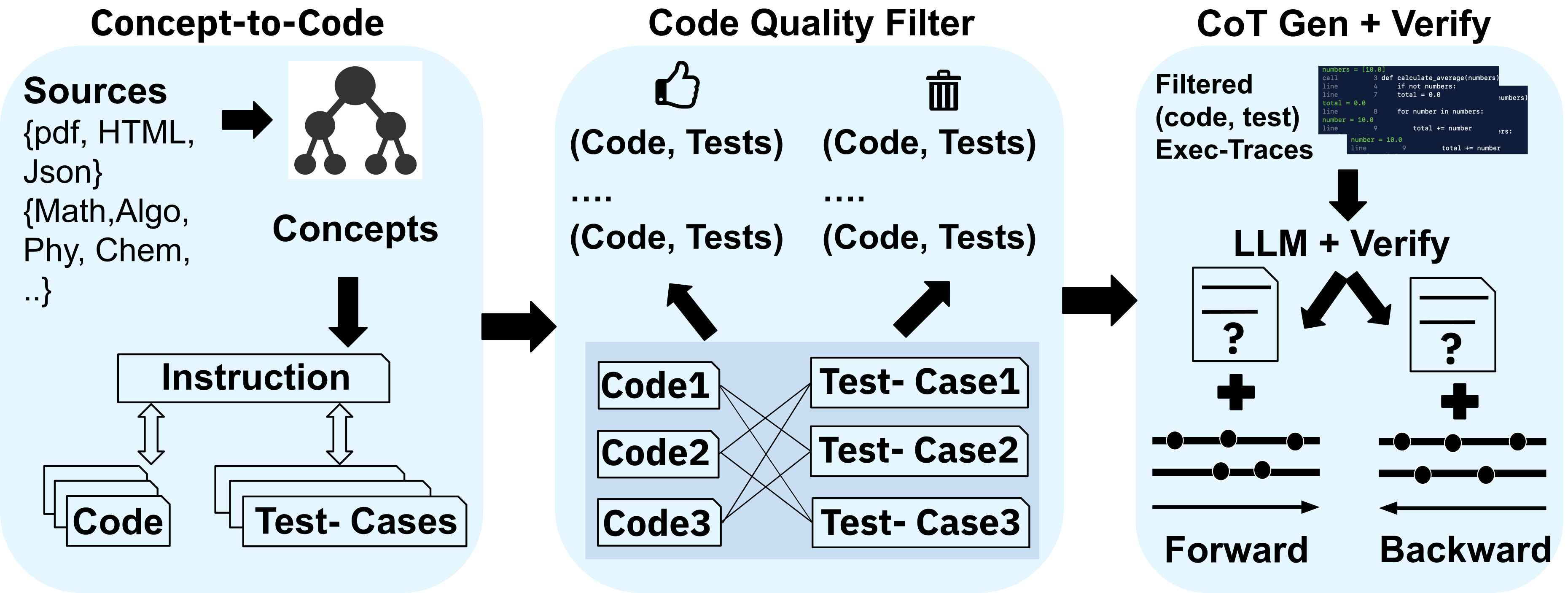}
        \caption{Overview of our three-stage data synthesis pipeline. \textbf{Stage A:} Generates candidate concepts, code, and tests from documents. \textbf{Stage B:} Uses Dual Agreement algorithm to identify highest-quality solution-test pairs. \textbf{Stage C:} Produces bi-directional training samples with trace-grounded \& verified rationales.}
        \label{fig:system_overview}
    \end{figure*}

    \begin{table}[t!]
    \centering\tiny
    \setlength{\tabcolsep}{3pt}
    \caption{Comparison of features in related works. Our work is the first to systematically combine execution-grounded verification with bi-directional CoT generation at scale. Abbreviations: NL=Natural Language.}
    \label{tab:related_work_comparison}
    \begin{tabular}{@{}l|ccccc@{}}
    \toprule
    \textbf{Work} & \textbf{Verifies} & \textbf{NL} & \textbf{Bidirectional} & \textbf{Verifies} & \textbf{Scalable} \\
    & \textbf{CoT} & \textbf{CoT} & \textbf{CoT} & \textbf{Output} & \textbf{Data} \\
    \midrule
    REVTHINK~\cite{chen2024reverse} & \xmark & \cmark & \cmark & \xmark & \cmark \\
    TRACED~\cite{ding2024traced} & \xmark & \xmark & \xmark & \cmark & \cmark \\
    CodeI/O~\cite{li2025codeio} & \xmark & \cmark & \cmark & \cmark & \cmark \\
    rStar-Coder~\cite{liu2025rstarcoderscalingcompetitivecode} & \xmark & \cmark & \xmark & \cmark & \cmark \\
    Jung et al.~\cite{jung2025codeexecutiongroundedsupervision} & \cmark & \cmark & \xmark & \cmark & \xmark \\
    SemCoder~\cite{ding2024semcodertrainingcodelanguage} & \xmark & \cmark & \cmark & \cmark & \cmark \\
    \midrule
    \textbf{Our Work} & \textbf{\gcmark} & \textbf{\gcmark} & \textbf{\gcmark} & \textbf{\gcmark} & \textbf{\gcmark} \\
    \bottomrule
    \end{tabular}%
    \end{table}

    Our contributions are:
    \begin{enumerate}[nosep]
        \item An end-to-end pipeline that generates execution-grounded CoT rationales and verifies each against the trace, demonstrated at scale with 54,000 examples.
        \item Systematic bi-directional CoT generation (forward and backward reasoning) enabling comprehensive code understanding.
        \item Empirical evidence that verified data quality outperforms quantity, and that execution grounding drives the largest gains in reasoning.
        \item Complete synthesis pipeline available as open-source contribution to support reproducibility.
    \end{enumerate}

    \section{Related Work}

    Our work builds upon code reasoning, execution-based program understanding, and synthetic data generation. Chain-of-Thought prompting~\cite{wei2022chain} and Program-of-Thought~\cite{chen2022program} lack systematic verification of reasoning steps. CodeI/O~\cite{li2025codeio} generates CoT rationales but does not verify intermediate reasoning against execution traces.

    Literature typically validates only final program outputs (Outcome Verification), offering no guarantee about intermediate reasoning. By verification in our work, we mean: validating each reasoning step against ground-truth execution traces---records of variable values, state transitions, and control flow. Methods like SemCoder~\cite{ding2024semcodertrainingcodelanguage} and rStar-Coder~\cite{liu2025rstarcoderscalingcompetitivecode} employ outcome
    verification but lack step-by-step reasoning validation. Recent work like TRACED~\cite{ding2024traced} and Execution Tuning~\cite{armengol2025execution} use execution traces for pre-training but do not generate natural language CoT with
    verification. Jung et al.~\cite{jung2025codeexecutiongroundedsupervision} generate
    CoT from traces using a teacher model but focus on single-direction reasoning and do not verify generated rationales against the trace. REVTHINK~\cite{chen2024reverse} explores bidirectional reasoning but without execution grounding or verification. Our setting is also distinct from inference-time execution-feedback methods~\cite{chen2023teaching,shinn2023reflexion}, which improve test-time behavior rather than grounding training data in observed execution.                                                                             
    We address these gaps by verifying intermediate reasoning against execution, supporting both reasoning directions, and generating trace-grounded CoT at scale.



    \section{Data Synthesis Pipeline}

    As shown in Figure~\ref{fig:system_overview}, our three-stage pipeline synthesizes code from diverse programming concepts, filtering high-quality code through execution-based consensus, and produce verified bi-directional CoT rationales at scale.

    \subsection{Stage A: Concept Sourcing and Synthesis}

    Rather than generating code from simple prompts, our pipeline builds a curriculum of programming concepts from high-quality sources.

    \textbf{Concept Extraction.} To ensure broad coverage, we extract programming concepts from permissively-licensed sources: StarCoder2-documentation~\cite{starcoder2-doc} and curated programming resources~\cite{freeprogrammingbooks}. Using a hybrid approach combining statistical NLP and LLM-based filtering, we retain concepts spanning data structures, algorithms, string manipulation, numerical computation, concurrency, and object-oriented patterns, balanced across difficulty levels from basic to advanced. Quality control via deduplication and scoring yields approximately 8,000 seed concepts.

    \textbf{Problem Synthesis.} For each concept, we synthesize complete problem artifacts through a five-step pipeline: (1) generate diverse natural language instructions across problem domains and difficulty levels; (2) produce formal signature skeletons to maintain consistency; (3) generate candidate solutions conforming to agreed interfaces; (4) identify test scenarios including edge cases; (5) generate unit tests with structural constraints for downstream parsing.

    Signatures from (2) enforce structural constraints on generated code (Figure~\ref{fig:signature-formats}), enabling function calls to be placed directly inside test assertions (Figure~\ref{fig:test-formats}). This consistent, parseable structure allows clean I/O and metadata extraction for downstream trace generation and verification.

    \subsection{Stage B: Consensus-Based Code Selection}

    Stage A generates candidate solutions and tests that may be incorrect or may have wrong expected outputs due to LLM generation errors. To identify high-quality code-test pairs, we adapt Dual Agreement~\cite{chen2022codetcodegenerationgenerated}, which leverages a key insight: \textit{if many independently generated solutions all pass the same independently generated tests, it is statistically unlikely that both the solutions and tests are
    incorrect.}

    We execute all $m \times n$ solution-test pairs in a sandboxed environment to construct a binary pass/fail matrix $M \in \{0,1\}^{m \times n}$, where $M[i, j] = 1$ if solution $i$ passes test $j$, and $0$ otherwise. We then partition solutions into clusters where all solutions within a cluster $C_i$ have identical pass/fail patterns across all tests. For each cluster, we assign a quality score:
    \[
    \text{Score}(C_i) = |C_i| \times |T_p(C_i)|
    \]
    where $|C_i|$ is cluster size and $|T_p(C_i)|$ is the number of tests all solutions pass. We select the highest-scoring cluster's canonical solution (shortest and most readable) and its corresponding test suite for downstream CoT generation. See Appendix~\ref{lst:dual-agreement-sample} for a concrete example showing how 5 solutions and 25 tests are clustered: Cluster 1 with 4 solutions passing 18 tests (score 72.0, selected) versus Cluster 2 with 1 solution passing 5 tests (score 5.0, rejected).


    \subsection{Stage C: Execution-Grounded CoT Generation \& Verification}

    For each solution-test pair from Stage B, we generate training samples by translating execution traces into natural language CoT rationales, then verifying each against the trace.

    \textbf{Trace Generation.} We use \texttt{pysnooper}~\cite{pysnooper} to instrument each executable solution, capturing function entry, each executed line, variable assignments with state transitions, and return values. This produces a ground-truth record of the program's operational semantics.

    \textbf{Trace Sanitization.} Raw pysnooper traces contain formatting artifacts (ANSI codes, timestamps, file paths) unsuitable for model consumption. We apply systematic sanitization via regex-based cleaning (see Appendix~\ref{lst:trace-to-cot-sample}), producing clean traces while preserving semantically meaningful transitions. This achieves approximately 40\% trace length reduction on average.

    \textbf{Question and CoT Generation.} We extract ground-truth I/O from sanitized traces and generate paired questions: forward (\texttt{``Given input X, what does the function return?''}) and backward (\texttt{``What input would cause output Y?''}). We prompt an LLM to narrate the execution trace for forward CoT and perform deductive reasoning from output state for backward CoT. Figure~\ref{fig:cot-templates} illustrates forward-only and backward-only examples. Notably, the model \textit{translates} factual execution rather than generating from scratch.

    \textbf{Verification Through Grounding.} Since CoT sentences do not map one-to-one to trace lines, we verify each rationale using a sliding window algorithm. For each sentence, we extract verifiable entities (variable names, values, control flow keywords) using an LLM, then check each entity against a lookahead window in the trace. We also string-match the predicted I/O in the rationale against ground-truth I/O from the test case. Rationales failing either check are discarded (see Appendix for algorithm details and rejection examples).

    \textbf{Dataset Assembly.} Verified rationales are assembled into three versions: forward-only and backward-only (Figure~\ref{fig:cot-templates}), and bi-directional (exemplified in Appendix Figure~\ref{lst:bidirectional-hard-sample}). Each sample contains a prompt section (blue tags in examples: instruction, code,
    and question(s)) and a response section (green regions in examples: trace-grounded CoT(s) and predicted I/O).

    \textbf{Language Scope and Transferability.} Our current implementation targets Python because it offers a convenient tracing and instrumentation layer through
    tools such as \texttt{pysnooper}. However, most of the pipeline is language-agnostic: concept sourcing, code/test synthesis, dual-agreement filtering, trace
    sanitization, rationale verification, and bi-directional data assembly transfer unchanged across languages. The language-specific component is trace acquisition
    together with extraction of verifiable entities such as variable values, state transitions, and executed control-flow decisions. Extending the method to another
    language therefore mainly requires replacing the tracing layer and adapting the parser for language-specific execution artifacts. In a language such as Java, this
    introduces additional engineering considerations including compilation filtering, maintaining shared code-test signatures, and scalable trace collection, but these
    changes are localized to the execution interface rather than the overall pipeline design.

\definecolor{promptblue}{RGB}{0, 102, 204}
\definecolor{lightgray}{RGB}{245, 245, 245}
\definecolor{lightblue}{RGB}{240, 248, 255}
\definecolor{lightgreen}{RGB}{240, 255, 240}
\definecolor{darkgray}{RGB}{64, 64, 64}
\definecolor{commentgreen}{RGB}{0, 128, 0}
\definecolor{keywordblue}{RGB}{0, 0, 255}
\definecolor{stringred}{RGB}{163, 21, 21}

\begin{figure}[t!]
  \centering
  \scriptsize
  \begin{tabular}{@{}p{0.48\columnwidth}@{\hspace{0.01\columnwidth}|}@{\hspace{0.01\columnwidth}}p{0.48\columnwidth}@{}}
  \toprule
  \textit{Forward CoT} (input$\rightarrow$output) & \textit{Backward CoT} (output$\rightarrow$input) \\
  \midrule
  \colorbox{lightblue}{\parbox{0.95\linewidth}{\scriptsize
  {\color{promptblue}\textbf{<Instruction>}}\\
  Given a weighted graph, find the most cost-effective path between two nodes.\\[0.1cm]

  {\color{promptblue}\textbf{<Function>}}\\
  \ttfamily def find\_cheapest\_path(graph, start, end): ...\\[0.1cm]
  \normalfont

  {\color{promptblue}\textbf{<Question>}}\\
  For the given network, what is the minimum cost and route from node 'A' to 'F'?
  }}
  &
  \colorbox{lightblue}{\parbox{0.95\linewidth}{\scriptsize
  {\color{promptblue}\textbf{<Instruction>}}\\
  Given a weighted graph, find a starting node for a cost-effective path.\\[0.1cm]

  {\color{promptblue}\textbf{<Function>}}\\
  \ttfamily def find\_cheapest\_path(graph, start, end): ...\\[0.1cm]
  \normalfont

  {\color{promptblue}\textbf{<Question>}}\\
  What start node could result in a path to 'F' with total cost of 7?
  }}
  \\[0.15cm]
  \colorbox{lightgreen}{\parbox{0.95\linewidth}{\scriptsize
  {\color{promptblue}\textbf{<CoT>}}
  \begin{enumerate}[leftmargin=8pt, itemsep=0.5pt, parsep=0pt, topsep=2pt]
  \item Initialize distances: 'A' to 0, all others to $\infty$. Priority queue starts with ('A', 0).
  \item Extract 'A' from queue. Update distances for neighbors 'B' and 'C'; add to queue.
  \item Continue extracting minimum, updating neighbors until destination 'F' reached.
  \item Backtrack from predecessors to reconstruct path. Final cost is 7.
  \end{enumerate}
  \vspace{0.05cm}
  {\color{promptblue}\textbf{<Predicted Output>}}\\
  \ttfamily (['A', 'C', 'E', 'F'], 7)
  }}
  &
  \colorbox{lightgreen}{\parbox{0.95\linewidth}{\scriptsize
  {\color{promptblue}\textbf{<CoT>}}
  \begin{enumerate}[leftmargin=8pt, itemsep=0.5pt, parsep=0pt, topsep=2pt]
  \item The final path to 'F' has cost 7. To deduce start, trace predecessors backwards.
  \item The predecessor of 'F' in cost-7 path must be 'E'. Edge cost from 'E' to 'F' is 2, so cost to reach 'E' is 5.
  \item Continue backtracking through 'C', which confirms starting at 'A' achieves total cost of 7.
  \end{enumerate}
  \vspace{0.05cm}
  {\color{promptblue}\textbf{<Predicted Input>}}\\
  \ttfamily 'A'
  }}
  \\
  \bottomrule
  \end{tabular}
  \vspace{-0.1cm}
  \caption{Forward and Backward CoT data format examples. Blue tags denote prompt components (instruction, function, question); green regions show model's trace-grounded reasoning and prediction. Code appears in \texttt{monospace}, reasoning in normal text.}
  \label{fig:cot-templates}
\end{figure}

\begin{figure}[t!]
  \centering
  \scriptsize
  \begin{tabular}{@{}m{0.48\columnwidth}@{\hspace{0.01\columnwidth}|}@{\hspace{0.01\columnwidth}}m{0.48\columnwidth}@{}}
  \toprule
  \textit{Function Signature} & \textit{Class Signature} \\
  \midrule
  \begin{minipage}[t]{0.46\columnwidth}
  \begin{lstlisting}[style=pythoncode, basicstyle=\ttfamily\tiny]
solution(freq_list: list[tuple[str,int]])
  -> encoding: dict[str, str]
\end{lstlisting}
\vspace{-0.1cm}
{\tiny\textbf{Extracted Metadata:}}\\
{\tiny
\textbullet~Function: \texttt{solution}\\
\textbullet~Param name: \texttt{freq\_list}\\
\textbullet~Param type: \texttt{list[tuple[str,int]]}\\
\textbullet~Return name: \texttt{encoding}\\
\textbullet~Return type: \texttt{dict[str,str]}}
  \end{minipage}
  &
  \begin{minipage}[t]{0.46\columnwidth}
  \begin{lstlisting}[style=pythoncode, basicstyle=\ttfamily\tiny]
Class HuffmanTree:
  __init__(self, freq_list: list[tuple[str,int]])
    -> None
  build_tree(self) -> tree: tuple
  get_encoding(self) -> codes: dict[str,str]
\end{lstlisting}
\vspace{-0.1cm}
{\tiny\textbf{Extracted Metadata:}}\\
{\tiny
\textbullet~Class: \texttt{HuffmanTree}\\
\textbullet~Constructor param: \texttt{freq\_list: list[tuple[str,int]]}\\
\textbullet~Method: \texttt{build\_tree}, returns \texttt{tree:tuple}\\
\textbullet~Method: \texttt{get\_encoding}, returns \texttt{codes:dict[str,str]}}
  \end{minipage}
  \\
  \bottomrule
  \end{tabular}
  \vspace{-0.1cm}
  \caption{Signature format templates with extracted metadata. Generated signatures specify function/class names, parameter names and types, and return variable names with types (e.g., \texttt{tree: tuple}, \texttt{codes: dict}). This metadata is stored to validate code/test consistency and prevent hallucinated names or type mismatches.}
  \label{fig:signature-formats}
\end{figure}

\begin{figure}[t!]
  \centering
  \scriptsize
  \begin{tabular}{@{}p{0.48\columnwidth}@{\hspace{0.01\columnwidth}|}@{\hspace{0.01\columnwidth}}p{0.48\columnwidth}@{}}
  \toprule
  \textit{Correct Format} \textcolor{correctgreen}{\cmark} & \textit{Prohibited Format} \textcolor{errorred}{\xmark} \\
  \midrule
  \begin{minipage}[t]{0.47\columnwidth}
  \colorbox{correctgreen!10}{\begin{minipage}{0.95\linewidth}
  \vspace{0.1cm}
  \texttt{\color{blue}def} \texttt{test\textunderscore basic():}\\
  \texttt{~~\color{green!60!black}\# Test basic case}\\
  \texttt{~~\color{blue}assert} \texttt{solution([1,2,3], 2) == [1]}
  \vspace{0.1cm}
  \end{minipage}}
  \end{minipage}
  &
  \begin{minipage}[t]{0.47\columnwidth}
  \colorbox{errorred!10}{\begin{minipage}{0.95\linewidth}
  \vspace{0.1cm}
  \texttt{\color{blue}def} \texttt{test\textunderscore wrong():}\\
  \texttt{~~\color{green!60!black}\# Variable outside assert}\\
  \texttt{~~lst = [1, 2, 3]}\\
  \texttt{~~\color{blue}assert} \texttt{solution(lst, 2) == [1]}
  \vspace{0.1cm}
  \end{minipage}}
  \end{minipage}
  \\
  \bottomrule
  \end{tabular}
  \vspace{-0.1cm}
  \caption{Test format requirements. Correct format (left) enables clean I/O extraction for trace generation with direct function calls in assert statements, while prohibited format (right) complicates trace analysis with intermediate variable assignments.}
  \label{fig:test-formats}
\end{figure}

    \begin{figure*}[t]
        \centering
        \includegraphics[width=1.4\columnwidth]{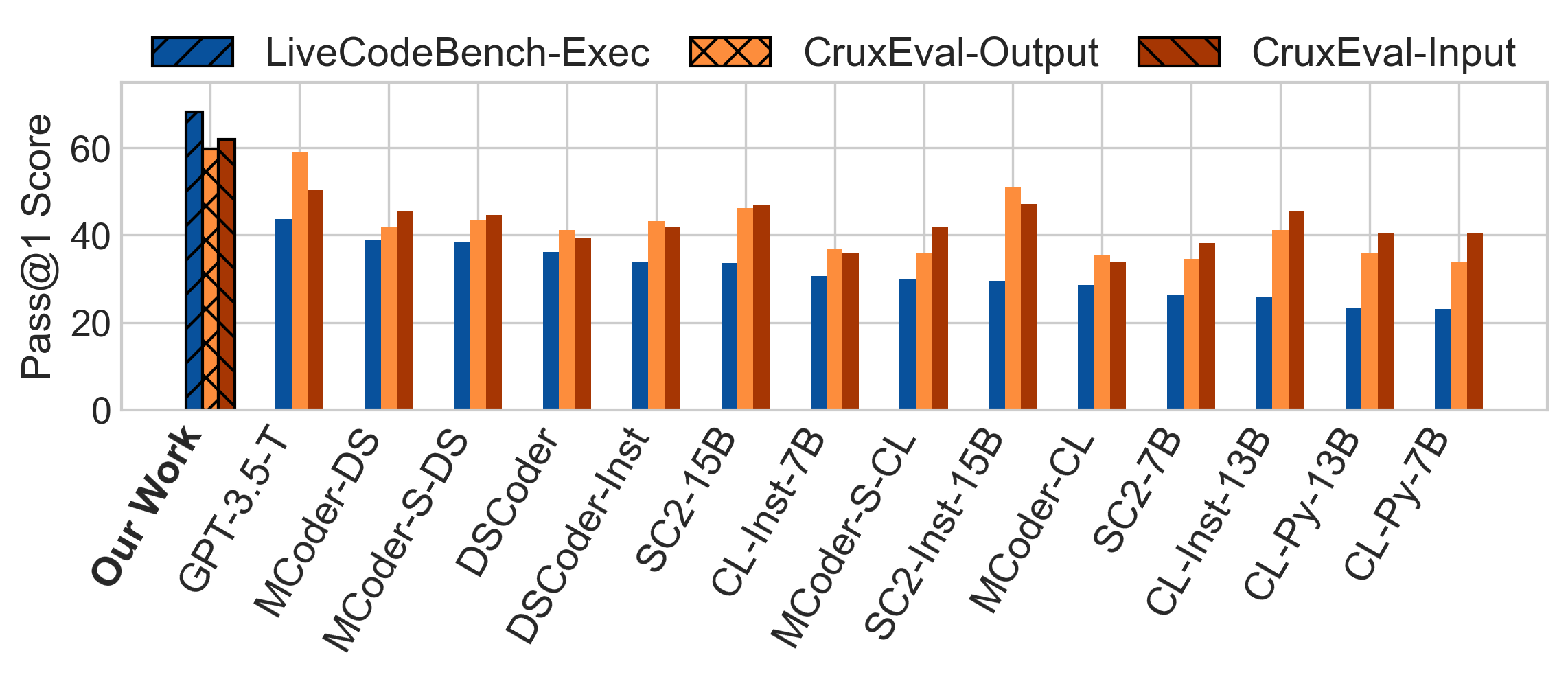}
        \caption{Comparison against published baselines~\cite{roziere2023codellama,li2023starcoder,guo2024deepseek,wei2023magicoder,ding2024semcodertrainingcodelanguage}. Our results have hatch pattern. CL=CodeLlama, SC2=StarCoder2, DS=DeepSeekCoder, MCoder=MagicCoder, S=SemCoder, Py=Python, Inst=Instruct.}
        \label{fig:sota-comparison}
        \vspace{-10pt}
    \end{figure*}

    \section{Experimental Setup}

    We use \texttt{Mistral-Small-24B-Instruct}~\cite{mistralsmall3} for all data synthesis (code, tests, signatures, and CoT narration from traces). We fine-tune \texttt{Granite-3.3-8B}~\cite{mishra2024granitecodemodelsfamily} and \texttt{Qwen2.5-Coder-7B}~\cite{hui2024qwen2} on our verified synthetic CoT data for 10 epochs with learning rate 2e-6, and report on LiveCodeBench-Exec (LCB)~\cite{jain2024livecodebench} and CruxEval~\cite{gu2024cruxeval} (Input/Output prediction) and HumanEval~\cite{chen2021evaluating}. All results are reported as Pass@$k$ (fraction of problems solved within $k$ attempts); we use $k{=}1$ unless otherwise stated. All evaluations use greedy decoding and default benchmark harness parameters.

    \textbf{Data Preparation.} Our pipeline yielded 15,000 deduplicated concepts and 85,000 functionally correct code-test pairs after Stage B. Our verification mechanism (Value, Transition, and Control Flow Matching) then filtered these, rejecting approximately 30\% of samples due to mismatches between generated CoTs and execution traces. Combined with deduplication and trace sanitization, this yielded 54,000 verified examples across three subsets: \textbf{Full (54k)}, \textbf{Model Answerability (25k)} keeps only problems that the base model fails to solve before fine-tuning, removing trivial or already likely memorized samples, and \textbf{Content-Rated Difficulty (18k)} retaining medium/hard problems.

    \textbf{Contamination Control.} We reduce benchmark overlap in three ways. Problems are synthesized from a concept curriculum derived from documentation and
    programming resources rather than benchmark prompts. The 25k \textbf{Model Answerability} subset described above further removes problems the base model already
    solves before fine-tuning. In addition, an LLM-based similarity screen on a representative random subset found no matches
    with LiveCodeBench or CruxEval under our overlap criteria. Together, these checks suggest that the synthesized training distribution is distinct from the
    evaluation sets.

    \textbf{Hyperparameter Selection.} We empirically validated our choice of 5 candidate solutions and 25 candidate tests per problem through grid search over configurations: solutions in $\{5, 10, 15, 20\}$ and tests in $\{5, 10, 20, 25, 30, 40, 50\}$. On a validation set of 500 problems with ground-truth solutions, quality (measured by alignment with ground truth) plateaus beyond 5 solutions and 25 tests, while computational cost scales linearly with matrix size. The 5$\times$25 configuration achieves $\sim$90\% of the maximum quality at 12.5\% of the cost of the 20$\times$50 configuration, representing the optimal quality-efficiency tradeoff (see Appendix Figure~\ref{fig:quality_cost_heatmap}). This choice is also consistent with the appendix validation of Dual Agreement Section~\ref{sec:dual_agreement_validation}, which shows a monotonic relationship between agreement score and empirical correctness on a larger held-out set. In practice, the execution matrix is parallelizable and cheaper than the LLM synthesis stages, making it a practical tradeoff rather than a bottleneck.

    \begin{table*}[t!]
        \centering
        \caption{Comprehensive evaluation results (Pass@$k$, $k \in \{1,5\}$ attempts per problem). Best per stage in \textbf{bold}; best overall in grey. All Pass@$k$ scores are percentages; blue subscripts show improvement over Base. Fwd: Forward only CoT, Bwd: Backward only CoT, Bi-Directional: Forward + Backward, FT: Fine-tuned.}
        \label{tab:comprehensive-results}
        \resizebox{\textwidth}{!}{%
        \begin{tabular}{C{2.5cm}|l|c|c|cc|cc}
        \hline
        \multirow{2}{*}{\textbf{Model}} & \multirow{2}{*}{\textbf{CoT Direc.}} & \multirow{2}{*}{\textbf{Data Subset}} &
        \textbf{LCB-Exec} &
        \multicolumn{2}{c|}{\textbf{CruxEval-O}} & \multicolumn{2}{c}{\textbf{CruxEval-I}} \\
        & & & \textbf{Pass@1} & \textbf{Pass@1} & \textbf{Pass@5} & \textbf{Pass@1} & \textbf{Pass@5} \\
            \hline
        \hline
        \parbox{2.5cm}{\centering Granite-3.3-8B-base} & Base (default) & N/A & 18.3 & 15.5 & 25.3 & 14.3 & 28.4 \\
        \hline
        \multirow{6}{*}{\parbox{2.5cm}{\centering Granite-3.3-8B (FT)}} & \multirow{3}{*}{Fwd} & 18k & 43.5$_{\textcolor{blue}{(+25.2)}}$ & 36.1$_{\textcolor{blue}{(+20.6)}}$ & 58.2$_{\textcolor{blue}{(+32.9)}}$ & 35.8$_{\textcolor{blue}{(+21.5)}}$ & 57.9$_{\textcolor{blue}{(+29.5)}}$ \\
        & & 25k (best) & \textbf{44.9}$_{\textcolor{blue}{(+26.6)}}$ & \textbf{42.7}$_{\textcolor{blue}{(+27.2)}}$ & \textbf{64.7}$_{\textcolor{blue}{(+39.4)}}$ & \textbf{40.2}$_{\textcolor{blue}{(+25.9)}}$ &
        \textbf{63.5}$_{\textcolor{blue}{(+35.1)}}$ \\
        & & 54k & 34.1$_{\textcolor{blue}{(+15.8)}}$ & 28.9$_{\textcolor{blue}{(+13.4)}}$ & 55.2$_{\textcolor{blue}{(+29.9)}}$ & 28.8$_{\textcolor{blue}{(+14.5)}}$ & 54.9$_{\textcolor{blue}{(+26.5)}}$ \\
        \cline{2-8}
        & Fwd & 25k & \cellcolor{highlightgray}{\textbf{44.9}$_{\textcolor{blue}{\textbf{(+26.6)}}}$} & 42.7$_{\textcolor{blue}{(+27.2)}}$ & 64.7$_{\textcolor{blue}{(+39.4)}}$ & 40.2$_{\textcolor{blue}{(+25.9)}}$ & 63.5$_{\textcolor{blue}{(+35.1)}}$ \\
        & Bwd & 25k & 35.4$_{\textcolor{blue}{(+17.1)}}$ & 39.3$_{\textcolor{blue}{(+23.8)}}$ & 61.3$_{\textcolor{blue}{(+36.0)}}$ & 41.5$_{\textcolor{blue}{(+27.2)}}$ & 64.8$_{\textcolor{blue}{(+36.4)}}$ \\
        & Bi-directional & 25k & 44.3$_{\textcolor{blue}{(+26.0)}}$ & \cellcolor{highlightgray}{\textbf{45.7}$_{\textcolor{blue}{\textbf{(+30.2)}}}$} & \cellcolor{highlightgray}{\textbf{67.4}$_{\textcolor{blue}{\textbf{(+42.1)}}}$} & \cellcolor{highlightgray}{\textbf{42.1}$_{\textcolor{blue}{\textbf{(+27.8)}}}$} & \cellcolor{highlightgray}{\textbf{65.2}$_{\textcolor{blue}{\textbf{(+36.8)}}}$} \\
        \hline
        \hline
        \parbox{2.5cm}{\centering Qwen2.5-Coder-7B} & Base (default) & N/A & 46.3 & 45.3 & 52.1 & 47.5 & 49.0 \\
        \hline
        \multirow{6}{*}{\parbox{2.5cm}{\centering Qwen2.5-Coder-7B (FT)}} & \multirow{3}{*}{Fwd} & 18k & 66.9$_{\textcolor{blue}{(+20.6)}}$ & 58.4$_{\textcolor{blue}{(+13.1)}}$ & 75.5$_{\textcolor{blue}{(+23.4)}}$ &
        57.2$_{\textcolor{blue}{(+9.7)}}$ & 69.4$_{\textcolor{blue}{(+20.4)}}$ \\
        & & 25k & \textbf{67.0}$_{\textcolor{blue}{(+20.7)}}$ & 57.5$_{\textcolor{blue}{(+12.2)}}$ & 73.9$_{\textcolor{blue}{(+21.8)}}$ & 59.8$_{\textcolor{blue}{(+12.3)}}$ & 67.5$_{\textcolor{blue}{(+18.5)}}$ \\
        & & 54k (best) & 66.5$_{\textcolor{blue}{(+20.2)}}$ & \textbf{58.6}$_{\textcolor{blue}{(+13.3)}}$ & \textbf{76.0}$_{\textcolor{blue}{(+23.9)}}$ & \textbf{60.5}$_{\textcolor{blue}{(+13.0)}}$ & \textbf{68.3}$_{\textcolor{blue}{(+19.3)}}$ \\
        \cline{2-8}
        & Fwd & 25k & 67.0$_{\textcolor{blue}{(+20.7)}}$ & 57.5$_{\textcolor{blue}{(+12.2)}}$ & 73.9$_{\textcolor{blue}{(+21.8)}}$ & 59.8$_{\textcolor{blue}{(+12.3)}}$ & 67.5$_{\textcolor{blue}{(+18.5)}}$ \\
        & Bwd & 25k & 57.5$_{\textcolor{blue}{(+11.2)}}$ & 50.4$_{\textcolor{blue}{(+5.1)}}$ & 69.8$_{\textcolor{blue}{(+17.7)}}$ & 61.2$_{\textcolor{blue}{(+13.7)}}$ & 69.1$_{\textcolor{blue}{(+20.1)}}$ \\
        & Bi-directional & 25k & \cellcolor{highlightgray}{\textbf{68.2}$_{\textcolor{blue}{\textbf{(+21.9)}}}$} & \cellcolor{highlightgray}{\textbf{59.7}$_{\textcolor{blue}{\textbf{(+14.4)}}}$} & \cellcolor{highlightgray}{\textbf{75.4}$_{\textcolor{blue}{\textbf{(+23.3)}}}$} & \cellcolor{highlightgray}{\textbf{61.9}$_{\textcolor{blue}{\textbf{(+14.4)}}}$} & \cellcolor{highlightgray}{\textbf{70.2}$_{\textcolor{blue}{\textbf{(+21.2)}}}$} \\

        \hline
        \end{tabular}%
        }
        \end{table*}

    \section{Results}
    We structure our evaluation as a sequential ablation: we first compare the three curation subsets (54k, 25k, 18k) to identify the best-performing training set, then use it to evaluate forward, backward, and bi-directional reasoning, and finally ablate the verification approach itself. Table~\ref{tab:comprehensive-results} presents all results.

    \subsection{Impact of Data Curation}
    We fine-tune models on three forward-only subsets to investigate data quality versus quantity.

    For \texttt{Granite-3.3-8B}, the \textbf{25k correctness-filtered dataset} substantially outperforms others. On LCB, it achieves 44.9\%, an absolute gain of \textbf{+26.6} over the base model, well ahead of the larger 54k set (34.1\%). This indicates that for complex reasoning, data verifiability is far more impactful than volume.

    For \texttt{Qwen2.5-Coder-7B}, all fine-tuned models outperform the baseline. While the 54k set edges ahead on CruxEval, the 25k dataset achieves the best LCB score (67.0\% Pass@1, \textbf{+20.7}) with comparable CruxEval performance. Given its stronger execution reasoning and smaller training cost, we selected the 25k set for subsequent experiments. Figure~\ref{fig:sota-comparison} compares our results against published baselines.

    \subsection{Impact of Reasoning Direction}

    Using the 25k dataset, we evaluated forward-only, backward-only, and bi-directional training. Table~\ref{tab:comprehensive-results} shows that bi-directional training improves both CruxEval and LiveCodeBench performance.

    For \texttt{Granite-3.3-8B}, the \textbf{bi-directional dataset} achieves the highest scores on CruxEval Output (45.7\% Pass@1, \textbf{+30.2}) and Input (42.1\% Pass@1, \textbf{+27.8}), while maintaining near-best LCB performance (44.3 vs.\ 44.9 for forward-only). This demonstrates that teaching both cause-to-effect and effect-to-cause reasoning yields complementary gains on understanding tasks without sacrificing execution performance.

    For \texttt{Qwen2.5-Coder-7B}, bi-directional training again achieves the best results on LiveCodeBench (68.2\%) and CruxEval Input (61.9\%), representing \textbf{+21.9} on LCB over the base model.

    \subsection{Impact of Verification Approach}

    We ablate verification methods using the 25k problem set with six conditions: (1) Base Model, (2) Few-Shot prompting, (3) Q\&A Only without CoT, (4) LLM-Generated CoT verified by an LLM judge (Mistral-Small-24B) without execution grounding, (5) Minimal Execution (1 solution, 1 test), and (6) Our Full Approach with Dual Agreement (Table~\ref{tab:verification_ablation}).  The \textit{LLM-Judge + LLM CoT} condition is our closest non-execution-grounded multi-candidate baseline. For each problem, we generate multiple CoT candidates and use a judge model to select the best rationale-and-answer pair. While this differs from majority-vote self-consistency, it captures the same intuition of improving quality through candidate selection without access to execution traces.

    \begin{table}[t]
    \centering
    \caption{Impact of verification approach (Pass@1). The table isolates the contribution of CoT supervision, execution grounding, and consensus-based artifact selection. All fine-tuned models use 25k bi-directional data. Tested on \texttt{Qwen2.5-Coder-7B}.}

    \label{tab:verification_ablation}
    \small
    \setlength{\tabcolsep}{3pt}
    \begin{tabular}{@{}lccc@{}}
    \toprule
    \textbf{Approach} & \textbf{LCB} & \textbf{CruxEval O/I} & \textbf{HumanEval} \\
    \midrule
    Base Model & 46.3 & 45.3/47.5 & 62.0 \\
    Few-Shot (3-shot) & 48.1 & 47.0/48.8 & 63.5 \\
    \midrule
    Q\&A Only (no CoT) & 52.7 & 50.2/51.3 & 69.0 \\
    \parbox[t]{2.8cm}{LLM-Judge\\+ LLM CoT} & 59.3 & 54.1/55.7 & 75.5 \\
    Min.\ Exec (1s, 1t) & 62.1 & 55.8/57.2 & 77.0 \\
    \midrule
    \textbf{Ours (Full)} & \textbf{68.2} & \textbf{59.7/61.9} & \textbf{81.5} \\
    \bottomrule
    \end{tabular}
    \end{table}


    The ablation reveals three key insights. First, execution grounding matters most---LLM-generated CoT without trace grounding (59.3 on LCB) lags well behind our full approach (68.2). Second, code correctness is a prerequisite: with only one solution and one test (Minimal Execution), there is no guarantee the code is correct, so the traces---and rationales grounded in them---may be unreliable. Our full approach uses Dual Agreement to ensure correctness before grounding, and the gap (62.1 vs.\ 68.2 on LCB) confirms this. This comparison isolates the value of consensus-based artifact selection from trace grounding alone. Minimal Execution provides trace-grounded rationales, but because it relies on only one solution and one test, it offers no consensus guarantee that the underlying code-test pair is correct. The improvement therefore reflects the combined value of grounding rationales in execution and grounding the execution itself in higher-confidence code-test artifacts selected by Dual Agreement. Third, CoT rationales help: removing them (Q\&A Only) drops performance. The gains also hold on HumanEval~\cite{chen2021evaluating} Pass@1 (62.0$\rightarrow$81.5\%), indicating that the benefit extends beyond input/output prediction to code generation quality. For reference, on CruxEval-I/O and LCB-Exec, our 7B model outperforms Scratchpad~\cite{nye2021scratchpad}, NExT~\cite{ni2024next}, and Concise Trace~\cite{ding2024semcodertrainingcodelanguage} approaches. SemCoder~\cite{ding2024semcodertrainingcodelanguage} reaches 61.8/63.5 on CruxEval-I/O but 58.5 on LCB-Exec (vs.\ our 68.2), using a different base model and training data. Our approach verifies each rationale against the trace before training, whereas SemCoder learns execution semantics without such verification.

    \subsection{Enhancing Instruction-Tuned Models}

\begin{figure}[t]
        \centering
        \includegraphics[width=\columnwidth]{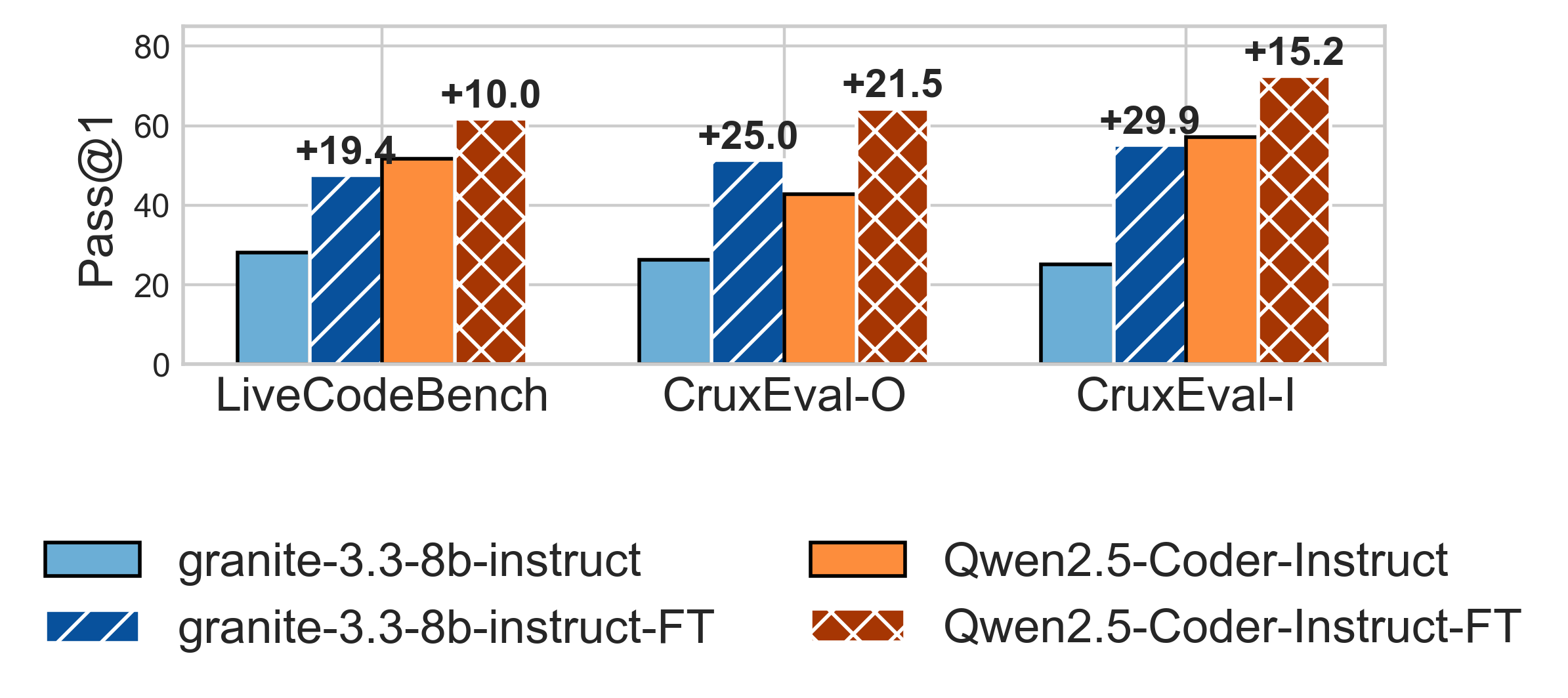}
        \caption{Performance boost from fine-tuning instruct models. Solid bars: baseline; hatched bars: improvement after fine-tuning with 25k bi-directional data.}
        \label{fig:instruct-boost}
        \vspace{-10pt}
    \end{figure}

    To confirm our execution-grounded data provides complementary benefits to instruction tuning, we also fine-tuned instruct-tuned versions of the models. As shown in Figure \ref{fig:instruct-boost}, this yields a \textbf{+39.9} gain on CruxEval Input for \texttt{Granite-3.3-8B-Instruct} and a \textbf{+21.5} gain on CruxEval Output for \texttt{Qwen2.5-Coder-7B-Instruct}.

    \subsection{CoT Quality Analysis}

    In addition to task accuracy, we ask whether execution-grounded training also improves the quality of the reasoning itself. We compare CoTs generated by fine-tuned \texttt{Qwen2.5-Coder-7B} against the baseline on CruxEval Output along two dimensions.

    \textbf{Consistency.} We measure how well reasoning steps align with the final answer. Our model shows a much stronger relationship between reasoning detail and logical correctness ($R^2 = 0.122$ vs.\ baseline $0.011$), and when its reasoning is internally consistent, it produces correct answers more reliably (AUC $= 0.567$ vs.\ $0.502$). In other words, our model's reasoning chains are not just longer---they are more logically coherent.

    \textbf{Information richness.} We measure how much technical detail each explanation contains using Shannon entropy over the distribution of unique technical tokens (variable names, concrete values, and control flow keywords) per explanation---higher entropy indicates more diverse, specific content. Compared to the base model, our fine-tuned model produces CoTs that are \textbf{761\% richer} in information content ($p < 0.001$). It references specific variable values, traces state transitions, and describes actual control flow from the execution---rather than generating generic summaries.

    \subsection{Deployment Footprint}
    The pipeline has been released as open source and integrated into downstream model
    post-training workflows. In particular, our 25k verified CoT subset was used in
    post-training stages of Granite-family of models~\cite{ibm2025granite4}, and the pipeline has also been
    adapted for use with standard open fine-tuning stacks such as OpenInstruct~\cite{olmo2025olmo3}. These integrations demonstrate practical downstream use as a reusable systems contribution that has already been exercised in real training pipelines.

    \section{Conclusion}
    We presented a method to generate factually grounded CoT training data at scale by grounding each reasoning step in execution traces. Results demonstrate that verified data quality drives improved reasoning, yielding peak gains of up to +22.2 on CruxEval, +26.6 on LiveCodeBench-Exec, and +19.5 on HumanEval over the corresponding base models we fine-tune. Beyond open-sourcing the pipeline, the resulting verified CoT data has already been used in post-training stages of deployed Granite-family models and integrated with existing fine-tuning stacks, indicating practical downstream utility. Future work includes extending the trace-acquisition layer to additional languages such as Java while preserving the same verification framework.

    \bibliography{sample-base}

    \appendix

\definecolor{promptblue}{RGB}{0, 102, 204}
\definecolor{darkgray}{RGB}{64, 64, 64}
\definecolor{commentgreen}{RGB}{0, 128, 0}
\definecolor{keywordblue}{RGB}{0, 0, 255}
\definecolor{stringred}{RGB}{163, 21, 21}
\definecolor{highlightyellow}{RGB}{255, 255, 204}
\definecolor{TraceVarState}{RGB}{0, 100, 0}
\definecolor{TraceTimestamp}{RGB}{120, 120, 120}
\definecolor{TraceAction}{RGB}{0, 0, 180}
\definecolor{TracePyNumber}{RGB}{128, 0, 128}
\definecolor{TracePyKeyword}{RGB}{0, 0, 180}

\lstdefinestyle{python-custom}{
    language=Python,
    basicstyle=\ttfamily\scriptsize,
    keywordstyle=\color{keywordblue},
    stringstyle=\color{stringred},
    commentstyle=\color{commentgreen},
    morecomment=[l]{\#},
    frame=tb, 
    framerule=0.5pt,
    numbers=none,
    breaklines=true,
    showstringspaces=false,
}

\newcounter{appendixfigure}
\renewcommand{\theappendixfigure}{A.\arabic{appendixfigure}}
\clearpage
\section*{Appendix}

\subsection*{Table of Contents}
\begin{itemize}[nosep, leftmargin=*]
    \item \textbf{Evaluation Setup} --- Benchmark descriptions and official links
    \item \textbf{Experimental Setup} --- Compute infrastructure and SFT hyperparameters
    \item \textbf{Dual Agreement Consensus Validation} --- Empirical validation and hyperparameter selection
    \item \textbf{Data Generation Prompts} --- Prompts for instruction, signature, code, test, and CoT narration
    \item \textbf{Dual Agreement Clustering Sample} --- Concrete pipeline example
    \item \textbf{Verifiable CoT Data Samples} --- Examples across reasoning directions and difficulty levels
    \item \textbf{Trace-to-CoT Mapping} --- End-to-end example from execution trace to bi-directional CoT
    \item \textbf{Sliding Window Verification} --- Algorithm details and rejection example
\end{itemize}

\vspace{0.5cm}

\section*{Evaluation Setup}
Our evaluation is conducted using two established, public benchmarks designed to test code reasoning and execution prediction capabilities. We provide direct links to the official benchmark pipelines to ensure full reproducibility of our results. For a fair and transparent comparison across all models, we use the default inference parameters provided by the CruxEval and LiveCodeBench evaluation harnesses without any model-specific tuning.

\begin{itemize}[nosep, leftmargin=*]
    \item \textbf{CruxEval:} A benchmark for code reasoning via input/output prediction. The official repository and data can be found at: \url{https://github.com/google-deepmind/cruxeval}
    
    \item \textbf{LiveCodeBench:} A benchmark featuring problems from live programming contests that tests code execution. The official repository is available at: \url{https://github.com/livecodebench/livecodebench}
\end{itemize}

\section*{Experimental Setup}
\subsubsection*{Compute Infrastructure}
We perform our experiments on a GPU cluster consisting of Dell XE9680 nodes. Each node has 96 CPU cores with 2 TB of system RAM. Each node is equipped with 8 Nvidia H100 GPUs, each having 80GB of GPU RAM. RHEL 9.4 is installed on the nodes. Each SFT run in our experiments uses 2 nodes (16 GPUs). Evaluation benchmarks are run on a single node using 8 GPUs.

\subsubsection*{SFT Setup}
We use Open Instruct Framework (\url{https://github.com/allenai/open-instruct}) for performing SFT. The hyper-parameters used for the training using 16 GPUs are described in table~\ref{tab-hyperparam}.

\begin{table}[h!]
  \centering
  \begin{tabular}{|c|c|}
  \hline
    Batch Size & 32 \\
    Per Device Batch Size & 1 \\
    Learning Rate & 2e-6 \\
    LR Scheduler & Linear \\
    Epochs & 10 \\
    Context Length & 8K \\
    \hline
  \end{tabular}
  \caption{SFT Hyper-parameters}
  \label{tab-hyperparam}
\end{table}

\section*{Dual Agreement Consensus Validation}
\label{sec:dual_agreement_validation}

\begin{figure}[h!]
    \centering
    \includegraphics[width=0.85\columnwidth]{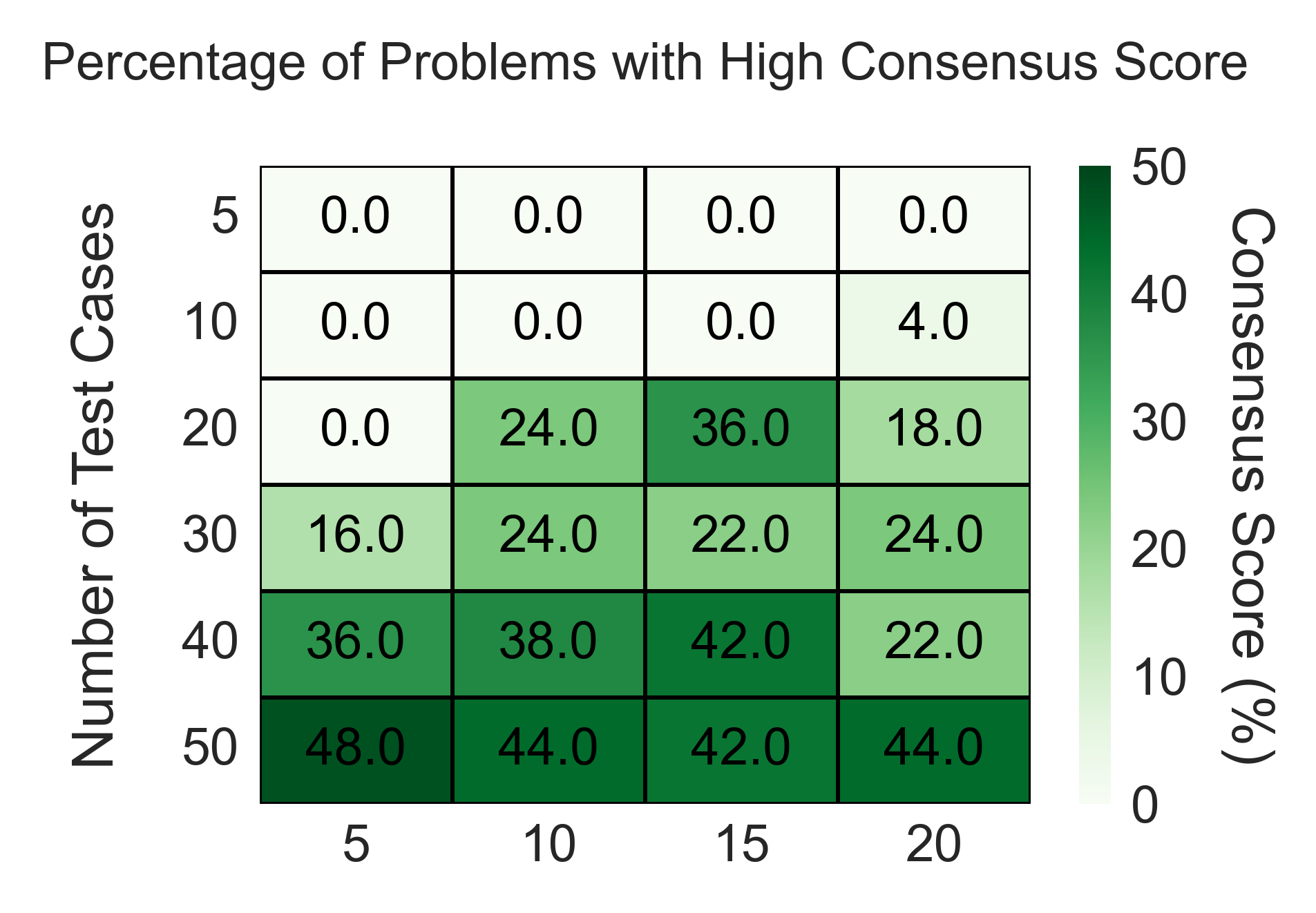}
    \caption{Percentage of problems achieving high consensus $\text{Score}(C_i)$ across configurations of candidate solutions ($m$, x-axis) and candidate tests ($n$, y-axis). Both dimensions contribute: $m$ strengthens cluster size $|C_i|$ while $n$ improves test coverage $|T_p(C_i)|$.}
    \label{fig:consensus_heatmap}
\end{figure}

We validate the Dual Agreement verification approach through a controlled study on 5,000 programming problems with ground-truth solutions. For each configuration of $m$ candidate solutions and $n$ candidate tests, we compute clusters $C_i$ and their quality scores $\text{Score}(C_i) = |C_i| \times |T_p(C_i)|$ as defined in Section 3.2, then measure the percentage of problems where the highest-scoring cluster achieves a high consensus score.

Figure~\ref{fig:consensus_heatmap} shows these consensus rates as a function of $m$ and $n$. The bottom-right region (high $m$, high $n$) achieves 70--90\% of problems with high consensus scores, while the upper-left (low $m$, low $n$) shows near-zero rates. This validates that \textit{both} dimensions are necessary: increasing $m$ provides more independent solutions for stronger cluster agreement ($|C_i|$), while increasing $n$ provides broader test coverage ($|T_p(C_i)|$), both of which contribute to higher $\text{Score}(C_i)$.

\subsection*{Score Formula Validation}

\begin{figure*}[h!]
    \centering
    \includegraphics[width=0.85\textwidth]{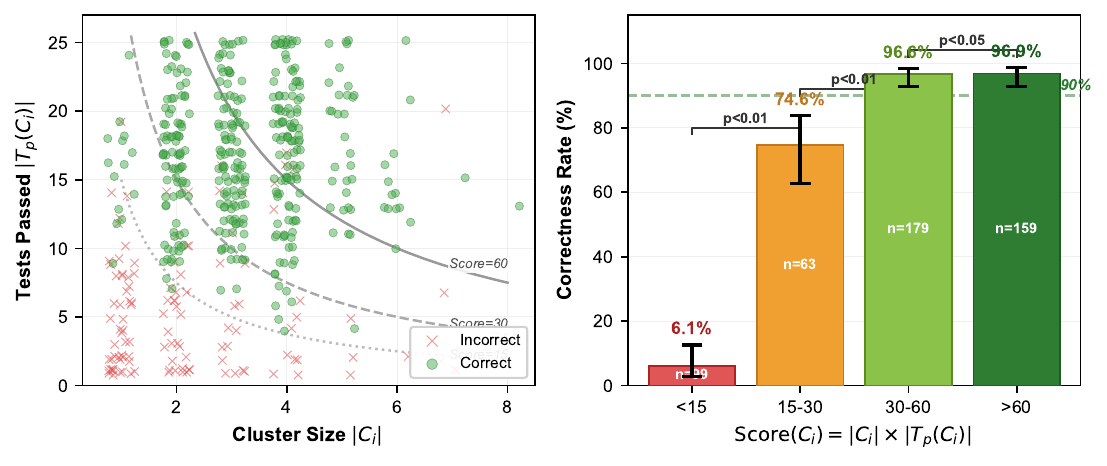}
    \caption{\textbf{Left:} Cluster size $|C_i|$ vs.\ tests passed $|T_p(C_i)|$ for the top-scoring cluster of each problem, colored by correctness. Iso-score curves show decision boundaries at $\text{Score} = 15, 30, 60$. \textbf{Right:} Correctness rate by score range with 95\% Wilson confidence intervals. Higher $\text{Score}(C_i) = |C_i| \times |T_p(C_i)|$ yields significantly higher correctness ($p < 0.01$), confirming the formula reliably selects correct solutions.}
    \label{fig:score_scatter}
\end{figure*}

\begin{figure*}[h!]
    \centering
    \includegraphics[width=0.85\textwidth]{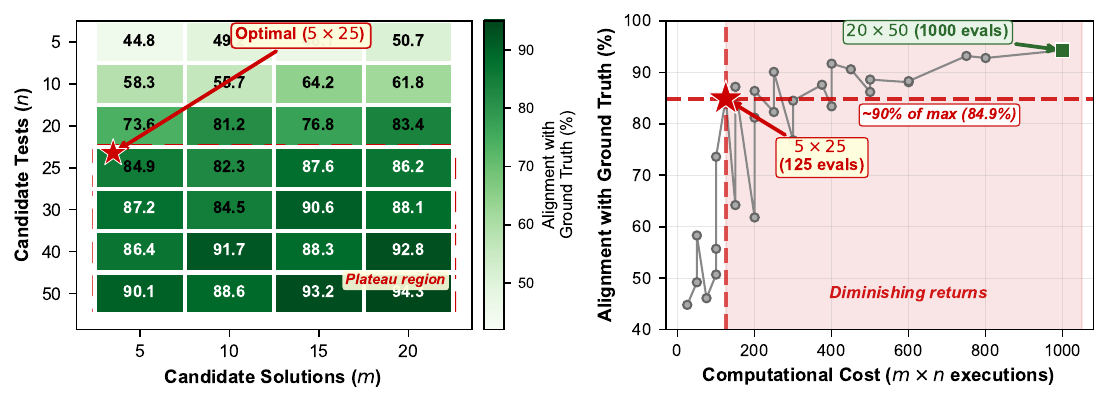}
    \caption{Hyperparameter selection for Dual Agreement. \textbf{Left:} Alignment with ground truth across configurations of candidate solutions ($m$) and tests ($n$). Quality plateaus beyond 25 tests. \textbf{Right:} Quality-cost tradeoff showing the $5 \times 25$ configuration achieves $\sim$90\% of maximum quality at 12.5\% of the computational cost.}
    \label{fig:quality_cost_heatmap}
\end{figure*}

Figure~\ref{fig:consensus_heatmap} shows that consensus improves with more candidates, but does the scoring formula $\text{Score}(C_i) = |C_i| \times |T_p(C_i)|$ actually distinguish correct solutions from incorrect ones? Figure~\ref{fig:score_scatter} answers this directly.

The \textbf{left panel} plots each problem's top cluster in the $|C_i|$ vs.\ $|T_p(C_i)|$ space, colored by whether the selected solution matches the ground truth. Correct solutions (green) concentrate in the upper-right region where both cluster size and test coverage are high, while incorrect solutions (red) scatter in the lower-left. The iso-score curves ($\text{Score} = 15, 30, 60$) show that the multiplicative structure of the formula naturally creates decision boundaries---problems above the $\text{Score} = 30$ curve are almost always correct.

The \textbf{right panel} quantifies this relationship by binning problems into score ranges and computing correctness rates with 95\% Wilson confidence intervals. The results show a clear monotonic relationship: problems with $\text{Score} < 15$ have only 6.1\% correctness, while those with $\text{Score} > 60$ achieve 96.9\% correctness ($p < 0.01$ between low and mid score ranges). The error bars confirm that these differences are statistically significant, not artifacts of sampling variance. This validates the formula's design: a solution must have \textit{both} strong agreement among independent implementations ($|C_i|$) \textit{and} broad test coverage ($|T_p(C_i)|$) to be selected, making it statistically unlikely that an incorrect solution survives.

\subsection*{Candidate Generation Hyperparameters: Quality-Cost Analysis}
\label{sec:hyperparam_quality_cost}

While Figure~\ref{fig:consensus_heatmap} above measures \textit{consensus}---whether independently generated solutions agree with each other---it does not directly tell us whether the selected solution is actually \textit{correct}. High consensus means many solutions behave identically, but they could all be wrong in the same way.

To address this, we conduct a separate analysis measuring \textbf{alignment with ground truth}: for each configuration of $m$ candidate solutions and $n$ candidate tests, we run Dual Agreement on a validation set of 500 problems with known-correct solutions, and measure how often the selected cluster's canonical solution matches the ground-truth answer.

Figure~\ref{fig:quality_cost_heatmap} presents this analysis in two panels:

\textbf{Left panel} shows alignment with ground truth as a heatmap across configurations. Unlike the consensus heatmap, which measures internal agreement, this measures external correctness. The key observations are: (1) quality generally increases with both the number of candidate solutions and tests, though with natural variance across individual cells; (2) a clear plateau emerges at approximately 25 tests---beyond this point, adding more tests yields marginal improvements regardless of the number of solutions; and (3) the $5 \times 25$ configuration (marked with a star) achieves 84.9\% alignment, which falls at the entry of the plateau region.

\textbf{Right panel} plots the same data as a quality-cost tradeoff curve, where cost is measured as the total number of solution-test executions ($m \times n$). This view makes the diminishing returns explicit: the $5 \times 25$ configuration requires only 125 executions and achieves approximately 90\% of the maximum quality obtained by the most expensive $20 \times 50$ configuration (1,000 executions). The shaded region highlights the zone of diminishing returns where substantially higher computational cost yields only marginal quality gains.

\textbf{In summary}, the two heatmaps are complementary: Figure~\ref{fig:consensus_heatmap} validates that Dual Agreement produces reliable consensus (an internal property), while Figure~\ref{fig:quality_cost_heatmap} confirms that this consensus translates to actual correctness (an external property) and identifies the optimal operating point at $5 \times 25$.

\section*{Data Generation Prompts}
This section provides the complete set of prompts used in our data synthesis pipeline. These prompts are referenced in Section 3 (Data Synthesis Pipeline) and are used throughout the hierarchical generation process in Stage A. The prompts are organized by their function in the pipeline, from instruction generation through solution and test case synthesis.

\subsection*{Instruction Generation Prompt}
This prompt generates diverse programming instructions for each concept, as described in Section 3.1.

\begin{tcolorbox}[
    enhanced, breakable,
    fontupper=\ttfamily\tiny,
    colback=blue!5!white,
    colframe=blue!20!white,
    boxrule=0.8pt, arc=3pt, boxsep=4pt,
    title={\textbf{instruction\_prompt}}, coltitle=black
]
\begin{lstlisting}[basicstyle=\ttfamily\tiny,breaklines=true]
You are an expert in Python programming and instructional design. Given the concepts and examples below, generate six distinct instructions with complexity level: {difficulty}. Ensure the tasks are as non-overlapping as possible while covering diverse aspects of the concepts.

COMPLEXITY GUIDELINES:
1. This {difficulty} difficulty should create {complexity_description}.
2. Solutions should span approximately {expected_lines} of code with rich, intricate logic maximized for 'hard' tasks.

INSTRUCTIONS FOR PROBLEM DIVERSITY:
1. Create problems that are fundamentally different in:
   - Problem domain: Include mathematics (e.g., algebra for equations and transformations, timing & durations for scheduling or sequencing, probabilities for statistical analysis, geometry for spatial computations), finance, data processing, algorithms, text processing, or system design. For 'hard' tasks, prioritize complex mathematical domains to ensure challenging synthesis.
   - Computational approach: Vary between iterative, recursive, dynamic programming, functional, or object-oriented programming.
   - Algorithmic complexity: Target specific time complexities (e.g., O(n) for easy, O(n log n) for medium, O(n^2) or higher for hard) appropriate to the difficulty, ensuring 'hard' tasks demand significant computational depth.
2. Before generating, analyze the concept's core principles and identify unique problem-solving strategies that leverage these principles, especially for mathematical domains in 'hard' tasks to maximize complexity and clarity.
3. Instructions may request either a standalone function named 'solution' or a class named 'Solution' with methods; indicate clearly if a class is required (e.g., 'implement a class') and specify the primary method name (e.g., 'compute') if applicable, otherwise assume 'compute' as the default primary method for classes.

Concept:
{concept}

Description:
{description}

Examples:
{examples}

Instructions should contain only the instruction text. Generate your response in the following format:
Instruction1:
{{}}

Instruction2:
{{}}

Instruction3:
{{}}

Instruction4:
{{}}

Instruction5:
{{}}

Instruction6:
{{}}
\end{lstlisting}
\end{tcolorbox}

\subsection*{Signature Generation Prompt}
This prompt analyzes instructions and generates appropriate function or class signatures, as described in Section 3.1.

\begin{tcolorbox}[
    enhanced, breakable,
    fontupper=\ttfamily\tiny,
    colback=blue!5!white,
    colframe=blue!20!white,
    boxrule=0.8pt, arc=3pt, boxsep=4pt,
    title={\textbf{signature\_prompt}}, coltitle=black
]
\begin{lstlisting}[basicstyle=\ttfamily\tiny,breaklines=true]
You are a Python programming expert. Given the instruction below, analyze how it should be implemented and provide the best signature skeleton. Follow these STRICT rules to determine the implementation type and format the output:
1. Decide if the instruction requires a standalone function or a class:
   - Choose a CLASS if the instruction EXPLICITLY says 'implement a class', 'create a class', or mentions methods like 'constructor', 'build_tree', etc., using the specified class name (e.g., 'HuffmanTree').
   - Otherwise, default to a standalone FUNCTION named 'solution'.
2. For a FUNCTION:
   - Format EXACTLY as: 'Function: name(param1: type1, param2: type2) -> return_type'
   - Include parameter names, types (infer if not specified), and return type (use 'unknown' if unclear).
3. For a CLASS:
   - Format EXACTLY as: 'Class: ClassName; __init__(self, param1: type1) -> return_type; method1(self, param2: type2) -> return_type; ...'
   - Use semicolons (;) to separate class name and methods.
   - Include '__init__' with parameters if implied, followed by all required methods.
   - Specify the primary method (named in instruction or 'compute' if unspecified) for testing.
   - Use 'unknown' for return types if not inferable.
4. RULES FOR FORMATTING:
   - Use ONLY spaces (no tabs, newlines, or escaped characters like '\').
   - Use EXACTLY the syntax shown (e.g., '__init__', '->', commas between params).
   - Do NOT add extra punctuation (e.g., colons after parentheses) or quotes around simple types (e.g., use 'Matrix', not '"Matrix"').
   - Do NOT deviate from the template---any variation is invalid.
   - Do NOT include explanations, prose, or multiple class definitions in one block---provide ONLY ONE signature skeleton.
   - Ensure the response is complete (no truncation) and matches the template EXACTLY.
5. Base your analysis ONLY on the instruction text, inferring types and outputs logically.
Instruction:
{instruction}

Return the signature skeleton INSIDE a code block, following the EXACT format below:
```text
Function: solution(input1: type1, input2: type2) -> return_type
```
or
```text
Class: ClassName; __init__(self, param1: type1) -> return_type; method1(self, param2: type2) -> return_type
```

Examples of CORRECT output:
```text
Function: solution(freq_list: list[tuple[str, int]]) -> dict[str, str]
```
```text
Class: HuffmanTree; __init__(self, freq_list: list[tuple[str, int]]) -> unknown; build_tree(self) -> tuple; get_encoding(self) -> dict[str, str]
```
Examples of INCORRECT output (DO NOT USE):
```text
Class: Matrix; **init**(self, data: list[list[int]]): add(self, other: "Matrix") -> "Matrix"
```
```text
Class: Polynomial; \ __init__(self, coeffs: list[float]) -> None; \ evaluate(self, value: float) -> float
```
```text
Class: Shape; area(self) -> float; Class: Circle; __init__(self, radius: float) -> None
```
Output MUST match the correct examples EXACTLY in format.
\end{lstlisting}
\end{tcolorbox}

\subsection*{Code Generation Prompts}
These prompts generate multiple candidate solutions for each instruction. We use separate prompts for function-based and class-based solutions.

\begin{tcolorbox}[
    enhanced, breakable,
    fontupper=\ttfamily\tiny,
    colback=ForestGreen!5!white,
    colframe=ForestGreen!20!white,
    boxrule=0.8pt, arc=3pt, boxsep=4pt,
    title={\textbf{Function Code Generation Prompt}}, coltitle=black
]
\begin{lstlisting}[basicstyle=\ttfamily\tiny,breaklines=true]
You are a Python programming expert. Given the instruction and signature details below, generate 5 functionally correct Python code adhering to these constraints:
1. **HIGH PRIORITY**: Implement a standalone function with name '{function_name}', inputs '{input_params}', and return type '{return_type}' EXACTLY as provided. Do NOT deviate from this signature.
2. Write all logic directly within '{function_name}'---do NOT define nested functions, even for multi-step problems; use variables or steps instead.
3. The function MUST ALWAYS RETURN A VALUE matching '{return_type}'.
4. Ensure the code is fully modular, self-contained, and does not rely on external code or global variables.
5. Optimize for readability, following Python best practices, with clear variable names and comments where necessary.
6. For hard difficulty, ensure the solution reflects the expected complexity: sophisticated long problems requiring complex algorithms and data structures (8-10 difficulty), spanning approximately 50-100+ lines with a difficulty score of 8-10 on a scale of 1-10.
7. **HIGH PRIORITY**: Generate EXACTLY FIVE distinct implementations, all strictly adhering to the provided signature:
   - Vary each implementation by:
     - Computational approach: Use distinct methods like iterative loops, recursion, dynamic programming, list comprehensions, or functional programming (e.g., map/filter/reduce), as appropriate to the instruction and difficulty.
     - Style: Alternate between verbose, step-by-step logic and concise, optimized solutions; use different commenting styles (e.g., inline vs. block comments).
     - Variable names: Use unique, meaningful names for variables and parameters in each response.
     - Complexity: Within the hard level, explore simpler vs. more intricate implementations (e.g., brute force vs. optimized algorithms).
   - Analyze the instruction to identify multiple viable strategies before generating solutions.
   - **ENSURE COMPLETENESS**: Each of the five implementations MUST be fully functional, including all required logic as specified in the instruction. Do NOT provide incomplete code (e.g., missing logic); generate all five implementations in full before terminating the response.
Instruction:
{instruction}

Signature Details:
- Function Name: {function_name}
- Inputs: {input_params}
- Return Type: {return_type}

Generate EXACTLY FIVE Python code blocks, all adhering to the provided signature, in this format:
```python
def {function_name}():  # Use the specified function name
    # Write the solution logic directly here (no nested functions)
    # Return the final output (MANDATORY)
    return ...  # Replace with actual value
```
\end{lstlisting}
\end{tcolorbox}

\begin{tcolorbox}[
    enhanced, breakable,
    fontupper=\ttfamily\tiny,
    colback=ForestGreen!5!white,
    colframe=ForestGreen!20!white,
    boxrule=0.8pt, arc=3pt, boxsep=4pt,
    title={\textbf{Class Code Generation Prompt}}, coltitle=black
]
\begin{lstlisting}[basicstyle=\ttfamily\tiny,breaklines=true]
You are a Python programming expert. Given the instruction and signature details below, generate 5 functionally correct Python code implementations adhering to these constraints:
1. HIGH PRIORITY: Implement a class with name '{class_name}' and methods as specified in '{method_signatures}' (including inputs and return types) EXACTLY as provided. Do NOT deviate from these signature details.
2. Include a constructor '{constructor_signature}' ONLY if explicitly provided in the signature details or if the instruction requires initialization of instance variables for the class to function correctly. Otherwise, omit the constructor.
3. Define the class with all necessary methods as specified, avoiding a function template.
4. Each method must be self-contained; each method MUST RETURN A VALUE matching its specified return type.
5. Ensure the code is fully modular, self-contained, and does not rely on external code or global variables.
6. Optimize for readability, following Python best practices, with clear variable names and comments where necessary.
7. For hard difficulty, ensure the solution reflects the expected complexity: sophisticated long problems requiring complex algorithms and data structures (8-10 difficulty), spanning approximately 50-100+ lines with a difficulty score of 8-10 on a scale of 1-10.
8. HIGH PRIORITY: Generate EXACTLY FIVE distinct implementations, all strictly adhering to the provided signature details:
   - Vary each implementation by:
     - Computational approach: Use distinct methods like iterative loops, recursion, dynamic programming, list comprehensions, or functional programming (e.g., map/filter/reduce), as appropriate to the instruction and difficulty.
     - Style: Alternate between verbose, step-by-step logic and concise, optimized solutions; use different commenting styles (e.g., inline vs. block comments).
     - Variable names: Use unique, meaningful names for variables and parameters in each response.
     - Complexity: Within the hard level, explore simpler vs. more intricate implementations (e.g., brute force vs. optimized algorithms).
   - Analyze the instruction to identify multiple viable strategies before generating solutions.
   - ENSURE COMPLETENESS: Each of the five implementations MUST be fully functional, including all required methods or logic as specified in the instruction. Do NOT provide incomplete code (e.g., missing method bodies or logic); generate all five implementations in full before terminating the response.
Instruction:
{instruction}

Signature Details:
- Class Name: {class_name}
- Constructor: {constructor_signature}
- Methods: {method_signatures}

Generate EXACTLY FIVE Python code blocks, all adhering to the provided signature details. Use this format when a constructor is needed:
```python
class {class_name}:  # Use the specified class name
    def __init__(self, freq_list):  # Constructor with specified parameters, only if required
        # Initialize attributes here
        pass
    def build_tree(self):  # Specified method
        # Construct the Huffman tree
        return ...  # Return as required
    def get_encoding(self):  # Specified method
        # Return the encoding dictionary (MANDATORY)
        return ...  # Replace with actual value
```
If no constructor is required, use this simpler format:
```python
class {class_name}:  # Use the specified class name
    def build_tree(self):  # Specified method
        # Construct the Huffman tree
        return ...  # Return as required
    def get_encoding(self):  # Specified method
        # Return the encoding dictionary (MANDATORY)
        return ...  # Replace with actual value
```
\end{lstlisting}
\end{tcolorbox}

\subsection*{Test Generation Prompts}
These prompts generate comprehensive unit test suites for validating the candidate solutions.

\begin{tcolorbox}[
    enhanced, breakable,
    fontupper=\ttfamily\tiny,
    colback=orange!5!white,
    colframe=orange!20!white,
    boxrule=0.8pt, arc=3pt, boxsep=4pt,
    title={\textbf{Test Scenario Identification Prompt}}, coltitle=black
]
\begin{lstlisting}[basicstyle=\ttfamily\tiny,breaklines=true]
You are an expert in Python testing and requirements analysis. Given the instruction and signature details below, analyze the task and identify a list of up to 10 concise test scenarios to guide test case generation. Each scenario must be a short hint (e.g., 'Test basic addition', 'Test empty input') to ensure all methods and key behaviors are tested, avoiding excessive detail. Focus on:
- Basic functionality of each method or function in the signature.
- Key behaviors or operations from the instruction.
- Broad coverage of the task's intent.
Return the list in this EXACT format, with no extra text outside the text block:
```text
Test scenario 1
Test scenario 2
...
```
Task Description:
{instruction}

Signature Details:
{signature_details}
\end{lstlisting}
\end{tcolorbox}

\begin{tcolorbox}[
    enhanced, breakable,
    fontupper=\ttfamily\tiny,
    colback=orange!5!white,
    colframe=orange!20!white,
    boxrule=0.8pt, arc=3pt, boxsep=4pt,
    title={\textbf{Function Test Generation Prompt}}, coltitle=black
]
\begin{lstlisting}[basicstyle=\ttfamily\tiny,breaklines=true]
You are an expert in Python testing and requirements analysis. Generate up to 10 isolated test cases for the following programming task based on the task description and the provided list of required test scenarios. Follow these CRITICAL GUIDELINES:
1. Each test case must be a standalone Python function (e.g., `def test_...():`), NOT defined within a class, to ensure easy parsing and execution.
2. Each test function must contain EXACTLY ONE assert statement.
3. Every assert statement MUST DIRECTLY call the function with specific inputs and compare its result to an expected value using a direct comparison (e.g., `==`, `is`, `in`, `!=`):
   - The solution to the task is a standalone function named '{function_name}', use `assert {function_name}(...) == ...` with all inputs packed into the call.
   - Do NOT:
     - Use variables or initializations outside the assert (e.g., `x = [1, 2]; assert {function_name}(x) == ...`).
     - Do NOT Use try-except blocks or check exceptions indirectly (e.g., `assert str(e) == ...`).
     - Do NOT use vague assertions (e.g., `assert == True`).
     - Use indirect comparisons (e.g., `.equals(...)`, timing checks).
     - Rely on external values; pack all necessary logic into the assert statement.
4. Generate up to 10 test cases, each corresponding to one of the required test scenarios provided below, ensuring each test directly calls the function with inputs matching the signature, all within the assert. If fewer than 10 scenarios are provided, generate only that number.
5. Verify that each test aligns with the task requirements, signature details, and the specified test scenario; all inputs must match the provided signature.
6. Ensure every assert statement is complete, specifying a concrete expected output value (e.g., a number, list, or string) and avoiding placeholders (e.g., '...'). Calculate the exact expected result based on the task description and signature for each test case.
Task Description:
{instruction}

Signature Details:
```python
{function_signature}
```

Required Test Scenarios:
{required_tests}

Use the template based on the signature (examples show dos and don'ts):
```python
# Do this:
def test_basic_functionality():
    # Test basic scenario
    assert {function_name}([1, 2, 3], 2) == 42

# Don't do this:
def test_basic_functionality_wrong():
    # Incorrect: variable outside assert
    lst = [1, 2, 3]
    assert {function_name}(lst, 2) == 42

# Don't do this:
def test_multi_assert_case():
    # Test scenario with multiple independent checks (not preferred)
    # Test Case 1
    assert {function_name}([1, 2], 1) == 10
    # Test Case 2
    assert {function_name}([3, 4], 1) == 20
```
\end{lstlisting}
\end{tcolorbox}

\begin{tcolorbox}[
    enhanced, breakable,
    fontupper=\ttfamily\tiny,
    colback=orange!5!white,
    colframe=orange!20!white,
    boxrule=0.8pt, arc=3pt, boxsep=4pt,
    title={\textbf{Class Test Generation Prompt}}, coltitle=black
]
\begin{lstlisting}[basicstyle=\ttfamily\tiny,breaklines=true]
You are an expert in Python testing and requirements analysis. Generate up to 10 isolated test cases for the following programming task based solely on the task description and the provided list of required test scenarios, without seeing the implementation. Follow these CRITICAL GUIDELINES:
1. Each test case must be a standalone Python function (e.g., `def test_...():`), NOT defined within a class, to ensure easy parsing and execution.
2. Each test function must contain EXACTLY ONE assert statement, unless the solution is a class with multiple methods and multiple asserts are needed to call logically connected methods (e.g., setup methods) before the primary method; in such cases, separate each assert with a numbered comment like `# Test Case 1`, `# Test Case 2`, etc., to distinguish them. For connected methods, prefer chaining them within a single assert statement (e.g., `{class_name}().setup(...).{primary_method}(...) == ...`) unless multiple asserts are unavoidable.
3. Every assert statement MUST DIRECTLY call the connected methods with specific inputs and compare its result to an expected value using a direct comparison (e.g., `==`, `is`, `in`, `!=`):
   - The solution to the task is a class named '{class_name}'. The primary method to test is '{primary_method}'. Instantiate it as `{class_name}()` and call methods directly in the assert; for logically connected methods, chain them within one assert (e.g., `assert {class_name}().method1(...).method2(...) == ...`). Do NOT:
     - Use variables or class instantiations outside the assert (e.g., `c = {class_name}(); assert c.method1(...).method2(...) == ...`).
     - Use try-except blocks or check exceptions indirectly (e.g., `assert str(e) == ...`).
     - Use vague assertions (e.g., `assert == True`).
     - Use indirect comparisons (e.g., `.equals(...)`, timing checks).
     - Rely on external values; pack all logic into the assert statement.
4. Generate up to 10 test cases, each corresponding to one of the required test scenarios provided below, ensuring each test directly calls the relevant method(s) with inputs matching their signature, all within the assert. If fewer than 10 scenarios are provided, generate only that number.
5. Verify that each test aligns with the task requirements, signature details, and the specified test scenario; all inputs must match the method signatures.
6. Ensure every assert statement is complete, specifying a concrete expected output value (e.g., a number, list, or string) and avoiding placeholders (e.g., '...'). Calculate the exact expected result based on the task description and signature for each test case.
Task Description:
{instruction}

Signature Details:
```python
Class: {class_name}
Class Methods:
{method_signatures}
Primary Method: {primary_method}
```

Required Test Scenarios:
{required_tests}

Generate test cases in this format, with each test in its own standalone function, using ONLY direct calls in asserts with complete expected values, packing all logic into the assert. Use the template based on the signature (examples show dos and don'ts):
- For class-based solutions:
```python
# Do this:
def test_basic_functionality():
    # Test basic scenario
    assert {class_name}().{primary_method}([1, 2, 3]) == 42

# Do this for logically connected methods, ensuring instantiation and calls are in one assert:
def test_connected_methods():
    # Test scenario where object instantiation and connected method calls are all in one assert statement
    assert {class_name}().setup([1, 2]).{primary_method}(3) == 42

# Don't do this:
def test_basic_functionality_wrong():
    # Incorrect: multiple asserts for class without logical connection
    assert {class_name}().{primary_method}([1, 2]) == 10
    assert {class_name}().{primary_method}([3, 4]) == 20

# Don't do this:
def test_setup_wrong():
    # Incorrect: setup outside assert
    obj = {class_name}()
    obj.setup([1, 2])
    assert obj.{primary_method}(3) == 42
```
\end{lstlisting}
\end{tcolorbox}

All custom analysis and evaluation scripts used to generate the results and figures presented in this paper, beyond the benchmark pipelines themselves, are included as part of the supplementary material.

\subsection*{Forward CoT Narration Prompt}
This prompt generates forward-reasoning Chain-of-Thought rationales by narrating how the program's execution transforms input to output using the execution trace, as described in Section 3.3.

\begin{tcolorbox}[
    enhanced, breakable,
    fontupper=\ttfamily\tiny,
    colback=blue!5!white,
    colframe=blue!20!white,
    boxrule=0.8pt, arc=3pt, boxsep=4pt,
    title={\textbf{forward\_cot\_narration\_prompt}}, coltitle=black
]
\begin{lstlisting}[basicstyle=\ttfamily\tiny,breaklines=true]
You are an expert in code execution who explains things like a patient teacher. Given a Python function, its raw execution trace for the given input and output, and provided input/output values, explain how the input leads to the output. Use the raw trace as evidence to guide your reasoning steps and ensure alignment and accuracy, presenting your explanation as a clear, deductive sequence of steps-starting with the code's structure and logic, then logically deducing the outcome. If there are loops, don't walk through each iteration; instead, summarize their overall effect and connections like a human would. Mention the predicted output naturally near the end of your steps, after describing the key operations, not at the start.

Code: ```python
{first_function}

Raw Trace: ```
{raw_trace}

Given Input: {input_val}
Given Output: {output_val}

Output Format:
<Steps>
1. [Natural first-person step]
2. [Next step]
</Steps>
{output}{output_val}{/output}
\end{lstlisting}
\end{tcolorbox}

\subsection*{Backward CoT Narration Prompt}
This prompt generates backward-reasoning Chain-of-Thought rationales by deducing the input from the output through reverse analysis of the execution trace, as described in Section 3.3.

\begin{tcolorbox}[
    enhanced, breakable,
    fontupper=\ttfamily\tiny,
    colback=blue!5!white,
    colframe=blue!20!white,
    boxrule=0.8pt, arc=3pt, boxsep=4pt,
    title={\textbf{backward\_cot\_narration\_prompt}}, coltitle=black
]
\begin{lstlisting}[basicstyle=\ttfamily\tiny,breaklines=true]
You are an expert in code execution, reasoning like a detective unraveling a mystery in reverse. Given a Python function, its raw execution trace, and the known output, deduce the exact input that produced the given output. Begin with the output and the final line of the code, working backward through the code and execution trace to reconstruct state transitions and variable changes leading to the input. Use the execution trace as evidence to ground your reasoning and ensure accuracy. Your explanation must be a clear, step-by-step deduction, starting with the output and reversing through the code's logic to deduce the input. Do not mention or assume any input until the final step, focusing each step on the code's operations, trace evidence, and logical conditions without jumping to conclusions. Use the trace to guide your steps, but do not rely on any input value upfront to shortcut the process. For loops or repetitive structures, summarize their overall effect and connections, as a human would reason. Your final deduced input must match the input recorded in the execution trace.

Code: ```python
{first_function}

Raw Trace: ```
{raw_trace}

Given Output: {output_val}

Output Format:
<Steps>
1. [Begin with the output and final code line, analyzing their relationship and trace evidence]
2. [Continue backward, examining prior code and trace to deduce state changes]
3. [Further reverse step, connecting logic and conditions]
...
N. [Conclude with the deduced input, tying together the reverse analysis]
</Steps>
{input_val}
\end{lstlisting}
\end{tcolorbox}

\subsection*{Appendix Sample: Dual Agreement Clustering Pipeline}
\label{lst:dual-agreement-sample}

This sample demonstrates Stage B (Consensus-Based Code Selection) output, showing
how Dual Agreement clusters solutions based on test consensus. The example traces
one task from synthesis through filtering to Stage C selection.

\vspace{6pt}
\textbf{Task Instruction:}
Write a Python function that computes the Greatest Common Divisor (GCD) of two
positive integers using the Euclidean algorithm. The function should efficiently
handle large numbers and return the GCD as an integer.

\vspace{6pt}
\textbf{Stage A Output (Input to Stage B):}
For this task instruction, Stage A synthesis produces:                                                                     
\begin{itemize}                                                                                                            
\item Candidate Solutions: 5 independently generated code implementations
\item Candidate Tests: 25 independently generated test cases
\end{itemize}
For this task instruction, Stage A synthesis produces \colorbox{yellow!40}{5 candidate solutions} and \colorbox{yellow!40}{25 candidate tests}.

\vspace{6pt}
\textbf{Stage B Processing: Execute All $5 \times 25 = 125$ Solution-Test Pairs}

After executing all pairs and grouping solutions by identical test outcomes,
we obtain multiple clusters. We show two representative clusters that emerge
from this process.

\vspace{12pt}
\textbf{Cluster 1 (High Consensus - \colorbox{green!40}{SELECTED for Stage C})}

Solutions in this cluster: 4 out of 5 generated implementations
Tests passing for all 4 solutions: 18 out of 25 generated tests

\vspace{6pt}

\vspace{6pt}
\begin{tcolorbox}[
    enhanced, breakable,
    fontupper=\ttfamily\scriptsize,
    colback=blue!5!white,
    colframe=blue!20!white,
    boxrule=0.8pt, arc=3pt, boxsep=4pt
]
\textbf{\color{blue!60!black}Representative Solutions from Cluster 1}

\textbf{Solution 1 (Iterative - Assignment-based):}
\begin{lstlisting}[style=pythoncode]
def solution(a, b):
    while b:
        a, b = b, a % b
    return a
\end{lstlisting}

\textbf{Solution 2 (Iterative - Temporary variable):}
\begin{lstlisting}[style=pythoncode]
def solution(a, b):
    while b != 0:
        temp = b
        b = a % b
        a = temp
    return a
\end{lstlisting}

\textbf{Solution 3 (Recursive):}
\begin{lstlisting}[style=pythoncode]
def solution(a, b):
    if b == 0:
        return a
    return solution(b, a % b)
\end{lstlisting}

\textbf{Solution 4 (Modulo with abs):}
\begin{lstlisting}[style=pythoncode]
def solution(a, b):
    a, b = abs(a), abs(b)
    while b:
        a, b = b, a % b
    return a
\end{lstlisting}

\medskip
\hrule
\medskip

\textbf{\color{blue!60!black}Tests All 4 Solutions Pass (18 out of 25, representative shown):}
\begin{lstlisting}[style=pythoncode]
def test_basic_case():
    assert solution(48, 18) == 6

def test_large_numbers():
    assert solution(1071, 462) == 21

def test_coprime_numbers():
    assert solution(17, 19) == 1

def test_multiple_common_factors():
    assert solution(360, 240) == 120
# ... 14 more tests covering edge cases,
#     large values, and reversed arguments
\end{lstlisting}

\end{tcolorbox}

\vspace{6pt}
\textbf{Cluster 1 Metrics:}
\begin{itemize}[nosep]
  \item \textbf{Solution Agreement:} 4 solutions with identical test pass patterns
  \item \textbf{Test Consensus:} 18 tests consistently passed by all 4 solutions
  \item \textbf{Quality Score:} $|C_i| \times |T_p(C_i)| = 4 \times 18 = \colorbox{yellow!40}{72.0}$
  \item \textbf{Interpretation:} High agreement across independent implementations suggests
        correctly generated code and valid test cases (CodeT finding)
\end{itemize}

\vspace{12pt}
\textbf{Cluster 2 (Low Consensus - \colorbox{red!40}{REJECTED})}

Solutions in this cluster: 1 out of 5 generated implementations
Tests passing for this solution: 5 out of 25 generated tests

\vspace{6pt}
\begin{tcolorbox}[
    enhanced, breakable,
    fontupper=\ttfamily\scriptsize,
    colback=blue!5!white,
    colframe=blue!20!white,
    boxrule=0.8pt, arc=3pt, boxsep=4pt
]
\textbf{\color{blue!60!black}Solution in Cluster 2}

\textbf{Solution (Incorrect - Off-by-one error):}
\begin{lstlisting}[style=pythoncode]
def solution(a, b):
    while b > 0:
        a, b = b, a % b
    return a + 1  # Bug
\end{lstlisting}

\medskip
\hrule
\medskip

\textbf{\color{blue!60!black}Tests This Solution Passes (5 out of 25):}
\begin{lstlisting}[style=pythoncode]
def test_zero_input():
    assert solution(0, 5) == 5

def test_same_values():
    assert solution(7, 7) == 7

def test_one_input():
    assert solution(1, 1) == 1

def test_power_of_two():
    assert solution(8, 4) == 4

def test_coprime_small():
    assert solution(3, 2) == 1
\end{lstlisting}

\end{tcolorbox}

\vspace{6pt}
\textbf{Cluster 2 Metrics:}
\begin{itemize}[nosep]
  \item \textbf{Solution Disagreement:} 1 solution with unique behavior (no agreement)
  \item \textbf{Low Test Consensus:} Only 5 tests, with failures indicating errors
  \item \textbf{Quality Score:} $|C_i| \times |T_p(C_i)| = 1 \times 5 = \colorbox{red!40}{5.0}$
  \item \textbf{Interpretation:} Low agreement suggests generation errors in either
        solutions or tests
\end{itemize}

\vspace{12pt}
\textbf{Dual Agreement Selection:}

Given 5 solutions and 25 tests (125 pairs executed), clustering reveals multiple
solution groups. Cluster 1 achieves \colorbox{yellow!40}{score 72.0} while
Cluster 2 achieves \colorbox{red!40}{score 5.0}. Following the CodeT methodology,
the highest-scoring cluster is selected, as it contains the most correctly
implemented solutions and valid test cases.

\vspace{6pt}
\textbf{Selected for Stage C:}
\begin{itemize}[nosep]
  \item \textbf{Canonical Solution:} Solution 1 from Cluster 1 (shortest, most readable)
  \item \textbf{Test Case:} Selected from 18 passing tests based on coverage of the canonical solution (see below)
\end{itemize}

\vspace{6pt}
\textbf{Test Case Selection.} The highest-scoring cluster contains multiple solutions and tests (here 4 solutions × 18 tests = 72 possible pairs). To avoid redundant training samples for the same coding task, we select one (solution, test) pair per problem. We pick the canonical solution and then rank the passing tests by estimated code coverage on that solution---favoring tests whose inputs exercise more branches, loops, and edge cases, while penalizing trivial or repeated inputs. We compared this coverage-based selection against an LLM-as-judge approach and found comparable downstream performance, so we use the simpler coverage-based criterion.

\vspace{6pt}
This pipeline demonstrates Dual Agreement's effectiveness: by clustering solutions
based on test consensus, it identifies the highest-quality code-test pairs, and coverage-based selection ensures each problem contributes one representative, high-coverage training sample to Stage C.


\section*{CoT Samples: Mapping an Execution Trace to Bi-Directional CoT}
We present representative examples of verifiable CoT data generated by our pipeline, covering forward, backward, and bi-directional reasoning on hard (Difficulty 4) problems. Each sample shows the user prompt and the model's trace-grounded response.

\label{lst:bidirectional-hard-sample}

This example walks through our complete pipeline on a single Difficulty 4 training sample involving multi-threaded parallel sorting. We show the problem instruction, canonical solution selected via Dual Agreement, test case with maximum coverage, pysnooper instrumentation, and the resulting execution trace that grounds the bi-directional CoT.

\begin{tcolorbox}[enhanced, breakable, colback=gray!5!white, colframe=gray!60!white, coltitle=black, colbacktitle=gray!30!white, boxrule=0.8pt, arc=3pt, boxsep=4pt,
    title={\textbf{Pipeline Inputs for Trace Generation}}]

\textbf{Problem Instruction (Difficulty 4):}
Implement a function \texttt{parallel\_sort} that takes a list of integers and a number of threads, divides the list into chunks, sorts each chunk in a separate thread, and merges the sorted chunks into a single sorted list.

\vspace{4pt}\hrule\vspace{4pt}

\textbf{Canonical Solution (selected by Dual Agreement):}
\begin{lstlisting}[style=python-custom]
import threading

def parallel_sort(numbers: list[int], num_threads: int) -> list[int]:
    if not numbers:
        return []
    chunk_size = len(numbers) // num_threads
    chunks = [numbers[i:i + chunk_size]
              for i in range(0, len(numbers), chunk_size)]
    threads = []
    sorted_chunks = [None] * num_threads
    lock = threading.Lock()

    def sort_chunk(index):
        sorted_chunks[index] = sorted(chunks[index])
        with lock:
            print(f"Chunk {index} sorted: {sorted_chunks[index]}")

    for i in range(num_threads):
        thread = threading.Thread(target=sort_chunk, args=(i,))
        threads.append(thread)
        thread.start()
    for thread in threads:
        thread.join()

    merged_list = []
    for chunk in sorted_chunks:
        if chunk:
            merged_list.extend(chunk)
    return sorted(merged_list)
\end{lstlisting}

\vspace{2pt}\hrule\vspace{4pt}

\textbf{Test Case (maximum coverage):}
\begin{lstlisting}[style=python-custom]
assert parallel_sort([-2, 3, -1, 0], 2) == [-2, -1, 0, 3]
\end{lstlisting}

\vspace{2pt}\hrule\vspace{4pt}

\textbf{Pysnooper Instrumentation:}
We instrument the canonical solution and run it with the test case:
\begin{lstlisting}[style=python-custom]
import pysnooper

@pysnooper.snoop()
def parallel_sort(numbers, num_threads):
    ...  # same function body as above

# Run with the test case
parallel_sort([-2, 3, -1, 0], 2)
\end{lstlisting}
The execution trace below is the resulting pysnooper output.
\end{tcolorbox}

\label{lst:trace-to-cot-sample}
\begin{tcolorbox}[
    enhanced, breakable,
    title={\textbf{Raw Execution Trace (pysnooper output)}},
    colback=gray!5!white,
    colframe=gray!60!white,
    coltitle=black, colbacktitle=gray!30!white,
    boxrule=0.8pt, arc=3pt, boxsep=4pt,
    fontupper=\ttfamily\scriptsize
]
\setlength{\parindent}{0pt} 
\textcolor{TraceVarState}{Starting var:..} numbers = [-2, 3, -1, 0] \\
\textcolor{TraceVarState}{Starting var:..} num\_threads = 2 \\
\textcolor{TraceTimestamp}{19:03:21.790438} \textcolor{TraceAction}{call} \phantom{xxxxxxxxx} \textcolor{TracePyNumber}{6} \textcolor{TracePyKeyword}{def} parallel\_sort(numbers: list[int], num\_threads: int) -> list[int]: \\
\textcolor{TraceTimestamp}{19:03:21.792564} \textcolor{TraceAction}{line} \phantom{xxxxxxxxxx} \textcolor{TracePyNumber}{7} \phantom{xx} \textcolor{TracePyKeyword}{if not} numbers: \\
\textcolor{TraceTimestamp}{19:03:21.793392} \textcolor{TraceAction}{line} \phantom{xxxxxxxxx} \textcolor{TracePyNumber}{11} \phantom{x} chunk\_size = len(numbers) // num\_threads \\
\textcolor{TraceVarState}{New var:.......} chunk\_size = 2 \\
\textcolor{TraceTimestamp}{19:03:21.793808} \textcolor{TraceAction}{line} \phantom{xxxxxxxxx} \textcolor{TracePyNumber}{12} \phantom{x} chunks = [numbers[i:i + chunk\_size] \textcolor{TracePyKeyword}{for} i \textcolor{TracePyKeyword}{in} range(0, len(numbers), chunk\_size)] \\
\textcolor{TraceVarState}{New var:.......} i = 0 \\
\textcolor{TraceTimestamp}{19:03:21.794953} \textcolor{TraceAction}{line} \phantom{xxxxxxxxx} \textcolor{TracePyNumber}{12} \phantom{x} chunks = [numbers[i:i + chunk\_size] \textcolor{TracePyKeyword}{for} i \textcolor{TracePyKeyword}{in} range(0, len(numbers), chunk\_size)] \\
\textcolor{TraceVarState}{Modified var:..} i = 2 \\
\textcolor{TraceTimestamp}{19:03:21.795618} \textcolor{TraceAction}{line} \phantom{xxxxxxxxx} \textcolor{TracePyNumber}{12} \phantom{x} chunks = [numbers[i:i + chunk\_size] \textcolor{TracePyKeyword}{for} i \textcolor{TracePyKeyword}{in} range(0, len(numbers), chunk\_size)] \\
\textcolor{TraceVarState}{New var:.......} chunks = [[-2, 3], [-1, 0]] \\
\textcolor{TraceTimestamp}{19:03:21.796810} \textcolor{TraceAction}{line} \phantom{xxxxxxxxx} \textcolor{TracePyNumber}{15} \phantom{x} threads = [] \\
\textcolor{TraceVarState}{New var:.......} threads = [] \\
\textcolor{TraceTimestamp}{19:03:21.798166} \textcolor{TraceAction}{line} \phantom{xxxxxxxxx} \textcolor{TracePyNumber}{16} \phantom{x} sorted\_chunks = [None] * num\_threads \\
\textcolor{TraceVarState}{New var:.......} sorted\_chunks = [None, None] \\
\textcolor{TraceTimestamp}{19:03:21.798746} \textcolor{TraceAction}{line} \phantom{xxxxxxxxx} \textcolor{TracePyNumber}{17} \phantom{x} lock = threading.Lock() \\
\textcolor{TraceVarState}{New var:.......} lock = <unlocked \_thread.lock object at 0x7fd060936f80> \\
\textcolor{TraceTimestamp}{19:03:21.799592} \textcolor{TraceAction}{line} \phantom{xxxxxxxxx} \textcolor{TracePyNumber}{19} \phantom{x} \textcolor{TracePyKeyword}{def} sort\_chunk(index): \\
\textcolor{TraceVarState}{New var:.......} sort\_chunk = <function parallel\_sort.<locals>.sort\_chunk at 0x7fd0607187c0> \\
\textcolor{TraceTimestamp}{19:03:21.800026} \textcolor{TraceAction}{line} \phantom{xxxxxxxxx} \textcolor{TracePyNumber}{24} \phantom{x} \textcolor{TracePyKeyword}{for} i \textcolor{TracePyKeyword}{in} range(num\_threads): \\
\textcolor{TraceVarState}{New var:.......} i = 0 \\
\textcolor{TraceTimestamp}{19:03:21.800171} \textcolor{TraceAction}{line} \phantom{xxxxxxxxx} \textcolor{TracePyNumber}{25} \phantom{xx} thread = threading.Thread(target=sort\_chunk, args=(i,)) \\
\textcolor{TraceVarState}{New var:.......} thread = <Thread(Thread-1 (sort\_chunk), initial)> \\
\textcolor{TraceTimestamp}{19:03:21.800504} \textcolor{TraceAction}{line} \phantom{xxxxxxxxx} \textcolor{TracePyNumber}{26} \phantom{xx} threads.append(thread) \\
\textcolor{TraceVarState}{Modified var:..} threads = [<Thread(Thread-1 (sort\_chunk), initial)>] \\
\textcolor{TraceTimestamp}{19:03:21.800979} \textcolor{TraceAction}{line} \phantom{xxxxxxxxx} \textcolor{TracePyNumber}{27} \phantom{xx} thread.start() \\
\textcolor{TraceVarState}{Modified var:..} threads = [<Thread(Thread-1 (sort\_chunk), stopped \dots)>] \\
\textcolor{TraceVarState}{Modified var:..} thread = <Thread(Thread-1 (sort\_chunk), stopped \dots)> \\
\textcolor{TraceVarState}{Modified var:..} sorted\_chunks = [[-2, 3], None] \\
\textcolor{TraceTimestamp}{19:03:21.801718} \textcolor{TraceAction}{line} \phantom{xxxxxxxxx} \textcolor{TracePyNumber}{24} \phantom{x} \textcolor{TracePyKeyword}{for} i \textcolor{TracePyKeyword}{in} range(num\_threads): \\
\textcolor{TraceVarState}{Modified var:..} i = 1 \\
\textcolor{TraceTimestamp}{19:03:21.802448} \textcolor{TraceAction}{line} \phantom{xxxxxxxxx} \textcolor{TracePyNumber}{25} \phantom{xx} thread = threading.Thread(target=sort\_chunk, args=(i,)) \\
\textcolor{TraceVarState}{New var:.......} thread = <Thread(Thread-2 (sort\_chunk), initial)> \\
\textcolor{TraceTimestamp}{19:03:21.803108} \textcolor{TraceAction}{line} \phantom{xxxxxxxxx} \textcolor{TracePyNumber}{26} \phantom{xx} threads.append(thread) \\
\textcolor{TraceVarState}{Modified var:..} threads = [<Thread(Thread-1 \dots)>, <Thread(Thread-2 \dots)>] \\
\textcolor{TraceTimestamp}{19:03:21.803716} \textcolor{TraceAction}{line} \phantom{xxxxxxxxx} \textcolor{TracePyNumber}{27} \phantom{xx} thread.start() \\
\textcolor{TraceVarState}{Modified var:..} threads = [<Thread(Thread-1 \dots)>, <Thread(Thread-2 \dots, stopped)>] \\
\textcolor{TraceVarState}{Modified var:..} thread = <Thread(Thread-2 (sort\_chunk), stopped \dots)> \\
\textcolor{TraceVarState}{Modified var:..} sorted\_chunks = [[-2, 3], [-1, 0]] \\
\textcolor{TraceTimestamp}{19:03:21.804575} \textcolor{TraceAction}{line} \phantom{xxxxxxxxx} \textcolor{TracePyNumber}{24} \phantom{x} \textcolor{TracePyKeyword}{for} i \textcolor{TracePyKeyword}{in} range(num\_threads): \\
\textcolor{TraceTimestamp}{19:03:21.804784} \textcolor{TraceAction}{line} \phantom{xxxxxxxxx} \textcolor{TracePyNumber}{29} \phantom{x} \textcolor{TracePyKeyword}{for} thread \textcolor{TracePyKeyword}{in} threads: \\
\textcolor{TraceVarState}{Modified var:..} thread = <Thread(Thread-1 (sort\_chunk), stopped \dots)> \\
\textcolor{TraceTimestamp}{19:03:21.804976} \textcolor{TraceAction}{line} \phantom{xxxxxxxxx} \textcolor{TracePyNumber}{30} \phantom{xx} thread.join() \\
\textcolor{TraceTimestamp}{19:03:21.805077} \textcolor{TraceAction}{line} \phantom{xxxxxxxxx} \textcolor{TracePyNumber}{29} \phantom{x} \textcolor{TracePyKeyword}{for} thread \textcolor{TracePyKeyword}{in} threads: \\
\textcolor{TraceVarState}{Modified var:..} thread = <Thread(Thread-2 (sort\_chunk), stopped \dots)> \\
\textcolor{TraceTimestamp}{19:03:21.805149} \textcolor{TraceAction}{line} \phantom{xxxxxxxxx} \textcolor{TracePyNumber}{30} \phantom{xx} thread.join() \\
\textcolor{TraceTimestamp}{19:03:21.805245} \textcolor{TraceAction}{line} \phantom{xxxxxxxxx} \textcolor{TracePyNumber}{29} \phantom{x} \textcolor{TracePyKeyword}{for} thread \textcolor{TracePyKeyword}{in} threads: \\
\textcolor{TraceTimestamp}{19:03:21.805299} \textcolor{TraceAction}{line} \phantom{xxxxxxxxx} \textcolor{TracePyNumber}{33} \phantom{x} merged\_list = [] \\
\textcolor{TraceVarState}{New var:.......} merged\_list = [] \\
\textcolor{TraceTimestamp}{19:03:21.805354} \textcolor{TraceAction}{line} \phantom{xxxxxxxxx} \textcolor{TracePyNumber}{34} \phantom{x} \textcolor{TracePyKeyword}{for} chunk \textcolor{TracePyKeyword}{in} sorted\_chunks: \\
\textcolor{TraceVarState}{New var:.......} chunk = [-2, 3] \\
\textcolor{TraceTimestamp}{19:03:21.807452} \textcolor{TraceAction}{line} \phantom{xxxxxxxxx} \textcolor{TracePyNumber}{35} \phantom{xx} \textcolor{TracePyKeyword}{if} chunk: \\
\textcolor{TraceTimestamp}{19:03:21.807554} \textcolor{TraceAction}{line} \phantom{xxxxxxxxx} \textcolor{TracePyNumber}{36} \phantom{xxx} merged\_list.extend(chunk) \\
\textcolor{TraceVarState}{Modified var:..} merged\_list = [-2, 3] \\
\textcolor{TraceTimestamp}{19:03:21.807622} \textcolor{TraceAction}{line} \phantom{xxxxxxxxx} \textcolor{TracePyNumber}{34} \phantom{x} \textcolor{TracePyKeyword}{for} chunk \textcolor{TracePyKeyword}{in} sorted\_chunks: \\
\textcolor{TraceVarState}{Modified var:..} chunk = [-1, 0] \\
\textcolor{TraceTimestamp}{19:03:21.807701} \textcolor{TraceAction}{line} \phantom{xxxxxxxxx} \textcolor{TracePyNumber}{35} \phantom{xx} \textcolor{TracePyKeyword}{if} chunk: \\
\textcolor{TraceTimestamp}{19:03:21.807792} \textcolor{TraceAction}{line} \phantom{xxxxxxxxx} \textcolor{TracePyNumber}{36} \phantom{xxx} merged\_list.extend(chunk) \\
\textcolor{TraceVarState}{Modified var:..} merged\_list = [-2, 3, -1, 0] \\
\textcolor{TraceTimestamp}{19:03:21.807846} \textcolor{TraceAction}{line} \phantom{xxxxxxxxx} \textcolor{TracePyNumber}{34} \phantom{x} \textcolor{TracePyKeyword}{for} chunk \textcolor{TracePyKeyword}{in} sorted\_chunks: \\
\textcolor{TraceTimestamp}{19:03:21.807932} \textcolor{TraceAction}{line} \phantom{xxxxxxxxx} \textcolor{TracePyNumber}{38} \phantom{x} \textcolor{TracePyKeyword}{return} sorted(merged\_list) \\
\textcolor{TraceTimestamp}{19:03:21.807997} \textcolor{TraceAction}{return} \phantom{xxxxx} \textcolor{TracePyNumber}{38} \phantom{x} \textcolor{TracePyKeyword}{return} sorted(merged\_list) \\
\textcolor{TraceVarState}{Return value:..} [-2, -1, 0, 3] \\
\textcolor{TraceTimestamp}{Elapsed time: 00:00:00.017813}
\end{tcolorbox}

\vspace{6pt}
From the execution trace above, we construct training samples in a \textbf{chat-style} format with alternating \texttt{User} and \texttt{Assistant} turns. The three reasoning directions differ in how these turns are arranged:

\begin{itemize}[nosep, leftmargin=*]
    \item \textbf{Forward-only:} \texttt{User} (instruction + code + input question) $\rightarrow$ \texttt{Assistant} (CoT + predicted output)
    \item \textbf{Backward-only:} \texttt{User} (instruction + code + output question) $\rightarrow$ \texttt{Assistant} (CoT + predicted input)
    \item \textbf{Bi-directional:} \texttt{User} (forward question) $\rightarrow$ \texttt{Assistant} (forward CoT) $\rightarrow$ \texttt{User} (backward question) $\rightarrow$ \texttt{Assistant} (backward CoT)
\end{itemize}

\vspace{4pt}
Below, we show the forward and backward turns for this example. A forward-only sample uses only the first turn; a backward-only sample uses only the second; a bi-directional sample concatenates both.

\vspace{6pt}
\begin{tcolorbox}[enhanced, breakable, colback=blue!5!white, colframe=blue!20!white, boxrule=0.8pt, arc=3pt, boxsep=4pt]
\textbf{\color{blue!60!black}User Prompt (Forward)}
\vspace{2pt}

The `parallel\_sort` function sorts a list of numbers by dividing it into chunks, sorting each chunk in a separate thread, and then merging the sorted chunks into a single sorted list using a lock to ensure thread-safe operations.
\vspace{2pt}

Here's the code:
\begin{lstlisting}[style=python-custom]

import threading

def parallel_sort(numbers: list[int], num_threads: int) -> list[int]:
    if not numbers:
        return []

    # Divide the list into chunks
    chunk_size = len(numbers) // num_threads
    chunks = [numbers[i:i + chunk_size] for i in range(0, len(numbers), chunk_size)]

    # Sort each chunk in a separate thread
    threads = []
    sorted_chunks = [None] * num_threads
    lock = threading.Lock()

    def sort_chunk(index):
        sorted_chunks[index] = sorted(chunks[index])
        with lock:
            print(f"Chunk {index} sorted: {sorted_chunks[index]}")

    for i in range(num_threads):
        thread = threading.Thread(target=sort_chunk, args=(i,))
        threads.append(thread)
        thread.start()

    for thread in threads:
        thread.join()

    # Merge the sorted chunks
    merged_list = []
    for chunk in sorted_chunks:
        if chunk:
            merged_list.extend(chunk)

    return sorted(merged_list)
\end{lstlisting}
\colorbox{highlightyellow}{\parbox{\dimexpr\linewidth-2\fboxsep}{\ttfamily\scriptsize
What output does the `parallel\_sort` function return when given the input `[-2, 3, -1, 0]` with 2 threads?
}}
\end{tcolorbox}
\begin{tcolorbox}[enhanced, breakable, colback=ForestGreen!5!white, colframe=ForestGreen!20!white, boxrule=0.8pt, arc=3pt, boxsep=4pt]
\textbf{\color{ForestGreen!60!black}Assistant Response }
\begin{lstlisting}[ frame=none, numbers=none, basicstyle={\ttfamily\scriptsize}, breaklines=true]
### Understand
The goal is to determine the output of the `parallel_sort` function when given the input `[-2, 3, -1, 0]` with 2 threads. The function is designed to sort a list of numbers using parallel processing by dividing the list into chunks, sorting each chunk in a separate thread, and then merging the sorted chunks into a final sorted list.

Key components of the function include:
- **Chunk Division**: The list is divided into chunks based on the number of threads.
- **Threaded Sorting**: Each chunk is sorted in a separate thread.
- **Merging**: The sorted chunks are merged and the final list is sorted before returning.

### Plan
1. **Chunk Division**: Calculate the size of each chunk and divide the list accordingly.
2. **Threaded Sorting**: Understand how each chunk is sorted in parallel.
3. **Merging**: Combine the sorted chunks and sort the final list.
4. **Verification**: Ensure the final sorted list is correct.

### Execute
1. **Chunk Division**:
   - The input list is `[-2, 3, -1, 0]` and the number of threads is `2`.
   - Calculate `chunk_size = len(numbers) // num_threads = 4 // 2 = 2`.
   - Divide the list into chunks: `chunks = [[-2, 3], [-1, 0]]`.

2. **Threaded Sorting**:
   - Each chunk is sorted in a separate thread:
     - Thread 1 sorts `[-2, 3]` resulting in `[-2, 3]`.
     - Thread 2 sorts `[-1, 0]` resulting in `[-1, 0]`.
   - The sorted chunks are stored in `sorted_chunks = [[-2, 3], [-1, 0]]`.

3. **Merging**:
   - Initialize `merged_list = []`.
   - Extend `merged_list` with each sorted chunk:
     - After adding `[-2, 3]`, `merged_list = [-2, 3]`.
     - After adding `[-1, 0]`, `merged_list = [-2, 3, -1, 0]`.
   - Sort the merged list: `sorted(merged_list) = [-2, -1, 0, 3]`.

4. **Verification**:
   - The final sorted list `[-2, -1, 0, 3]` is consistent with the expected behavior of sorting the entire list.

### Reflect
The reasoning process involved understanding how the list is divided, sorted in parallel, and merged. The final step of sorting the merged list ensures the entire list is sorted correctly. The approach was thorough, considering each step of the function's logic. The final output aligns with the expected result, confirming the correctness of the reasoning.
\end{lstlisting}
\colorbox{highlightyellow}{\parbox{\dimexpr\linewidth-2\fboxsep}{\ttfamily\scriptsize
Predicted Output: [-2, -1, 0, 3]
}}
\end{tcolorbox}

\vspace{6pt}
\begin{tcolorbox}[enhanced, breakable, colback=blue!5!white, colframe=blue!20!white, boxrule=0.8pt, arc=3pt, boxsep=4pt]
\textbf{\color{blue!60!black}User Prompt (Backward)}\\
\vspace{2pt}
\colorbox{highlightyellow}{\parbox{\dimexpr\linewidth-2\fboxsep}{\ttfamily\scriptsize
What input (list and number of threads) to `parallel\_sort` produces the output `[-2, -1, 0, 3]`?
}}
\end{tcolorbox}

\begin{tcolorbox}[enhanced, breakable, colback=ForestGreen!5!white, colframe=ForestGreen!20!white, boxrule=0.8pt, arc=3pt, boxsep=4pt]
\textbf{\color{ForestGreen!60!black}Assistant Response }
\begin{lstlisting}[ frame=none, numbers=none, basicstyle={\ttfamily\scriptsize}, breaklines=true]
### Understand
The function `parallel_sort` takes a list of integers and a number of threads. It divides the list into chunks, sorts each chunk in a separate thread, merges the sorted chunks into `merged_list`, and returns `sorted(merged_list)`. The output is `[-2, -1, 0, 3]`. To find the input, I need to work backward through each operation.

### Plan
1. Start from the output `[-2, -1, 0, 3]` and reverse the final `sorted()` call.
2. Reverse the merge to find the sorted chunks.
3. Reverse the per-chunk sorting to find the original chunks.
4. Reverse the chunk division to recover the input list and number of threads.

### Execute
1. **Reverse the final sort**: The function returns `sorted(merged_list) = [-2, -1, 0, 3]`. The `merged_list` before sorting must contain the same elements. Since merging extends chunks sequentially, `merged_list` was built by appending `[-2, 3]` followed by `[-1, 0]`, giving `merged_list = [-2, 3, -1, 0]`. Applying `sorted()` to this yields `[-2, -1, 0, 3]`, which matches.

2. **Reverse the merge**: The merge loop iterates over `sorted_chunks` and extends `merged_list`. Since `merged_list = [-2, 3, -1, 0]` and the first extension added `[-2, 3]` and the second added `[-1, 0]`, we get `sorted_chunks = [[-2, 3], [-1, 0]]`.

3. **Reverse the chunk sorting**: Each chunk was sorted individually by `sort_chunk`. For the first chunk, `sorted(chunk_0) = [-2, 3]`. Since `[-2, 3]` is already sorted, the original `chunk_0 = [-2, 3]`. For the second chunk, `sorted(chunk_1) = [-1, 0]`. Since `[-1, 0]` is already sorted, the original `chunk_1 = [-1, 0]`.

4. **Reverse the chunk division**: The chunks `[[-2, 3], [-1, 0]]` have size 2 each. `chunk_size = len(numbers) // num_threads`, so `2 = 4 // num_threads`, giving `num_threads = 2`. The original list is the concatenation of the chunks: `[-2, 3] + [-1, 0] = [-2, 3, -1, 0]`.

### Reflect
By reversing each stage---final sort, merge, chunk sorting, and chunk division---we recover the full input. The deduced `numbers = [-2, 3, -1, 0]` with `num_threads = 2` produces chunks `[[-2, 3], [-1, 0]]`, which sort to `[[-2, 3], [-1, 0]]`, merge to `[-2, 3, -1, 0]`, and `sorted([-2, 3, -1, 0])` returns `[-2, -1, 0, 3]`, confirming consistency.
\end{lstlisting}
\colorbox{highlightyellow}{\parbox{\dimexpr\linewidth-2\fboxsep}{\ttfamily\scriptsize
Predicted Input: [-2, 3, -1, 0], 2
}}
\end{tcolorbox}

\section*{Sliding Window-based CoT Verification}

Since natural language explanations do not map one-to-one to trace lines (e.g., one sentence may summarize several steps), we use a sliding window algorithm to verify each rationale against the trace. The verifier maintains a pointer to the current position in the execution trace. For each CoT sentence, we extract verifiable entities---variable names, values, and control flow keywords---using an LLM (extraction only, not reasoning). For each extracted entity, the algorithm scans a lookahead window of $K{=}15$ trace steps from the current pointer. A reasoning step is accepted if:

\begin{enumerate}[nosep]
    \item \textbf{Event Matching}: A trace step within the window contains the cited variable modification or value.
    \item \textbf{State Consistency}: The cited value exists in the program's accumulated state at the current pointer.
\end{enumerate}

If neither condition holds, the entity is ungrounded and the rationale is flagged. As a second check, the final answer stated in the rationale is string-matched against ground-truth I/O from the test case. Rationales failing either check are discarded. This process rejected approximately 30\% of generated rationales.

\subsection*{Detailed Rejection Example}

Below we show a full trace and rejected rationale side by side for \texttt{binary\_search([1,3,5,7], 5)}, followed by the verification walkthrough.

\noindent
\begin{minipage}[t]{0.48\textwidth}
\begin{tcolorbox}[enhanced, colback=gray!5!white, colframe=gray!60!white, coltitle=black, colbacktitle=gray!30!white, boxrule=0.8pt, arc=3pt, boxsep=4pt,
    title={\textbf{Full Trace}}]
\ttfamily\scriptsize
\setlength{\parindent}{0pt}
\textbf{[1]} \textcolor{TraceVarState}{Starting var:} arr=[1,3,5,7], target=5\\
\textbf{[2]} \textcolor{TraceAction}{line} lo = 0\\
\textbf{[3]} \textcolor{TraceVarState}{New var:} lo = 0\\
\textbf{[4]} \textcolor{TraceAction}{line} hi = len(arr) - 1\\
\textbf{[5]} \textcolor{TraceVarState}{New var:} hi = 3\\
\textbf{[6]} \textcolor{TraceAction}{line} while lo <= hi:\\
\textbf{[7]} \textcolor{TraceAction}{line} mid = (0+3)//2\\
\textbf{[8]} \textcolor{TraceVarState}{New var:} mid = 1\\
\textbf{[9]} \textcolor{TraceAction}{line} if arr[1]==5 $\to$ False\\
\textbf{[10]} \textcolor{TraceAction}{line} elif arr[1]<5 $\to$ True\\
\textbf{[11]} \textcolor{TraceAction}{line} lo = mid + 1\\
\textbf{[12]} \textcolor{TraceVarState}{Modified var:} lo = 2\\
\textbf{[13]} \textcolor{TraceAction}{line} while lo <= hi:\\
\textbf{[14]} \textcolor{TraceAction}{line} mid = (2+3)//2\\
\textbf{[15]} \textcolor{TraceVarState}{Modified var:} mid = 2\\
\textbf{[16]} \textcolor{TraceAction}{line} if arr[2]==5 $\to$ True\\
\textbf{[17]} \textcolor{TraceAction}{line} return mid\\
\textbf{[18]} \textcolor{TraceVarState}{Return value:} 2
\end{tcolorbox}
\end{minipage}%
\hfill
\begin{minipage}[t]{0.48\textwidth}
\begin{tcolorbox}[enhanced, colback=red!3!white, colframe=red!40!white, coltitle=black, colbacktitle=red!15!white, boxrule=0.8pt, arc=3pt, boxsep=4pt,
    title={\textbf{Rejected Rationale}}]
\begin{lstlisting}[frame=none, numbers=none, basicstyle={\ttfamily\scriptsize}, breaklines=true]
### Understand
The function performs binary search on a sorted array to find the index of a target value.

### Plan
1. Initialize lo=0, hi=len(arr)-1.
2. Compute mid, compare arr[mid] with target.
3. Narrow search range until found or exhausted.

### Execute
1. Start with lo=0 and hi=3.
2. First iteration: mid=(0+3)//2=1. arr[1]=3, which is less than target 5, so lo becomes 2.
3. Second iteration: mid=(2+3)//2=2. arr[2]=5. Since 5>5 is false and 5<5 is false, we enter the else branch and set hi=1.
4. Loop continues but lo>hi, so we exit.

### Reflect
The search narrowed correctly in the first iteration but failed to find the target in the second, returning -1.
\end{lstlisting}
\colorbox{highlightyellow}{\parbox{\dimexpr\linewidth-2\fboxsep}{\ttfamily\scriptsize
Predicted Output: -1
}}
\end{tcolorbox}
\end{minipage}

\begin{tcolorbox}[enhanced, breakable, colback=gray!5!white, colframe=gray!60!white, coltitle=black, colbacktitle=gray!30!white, boxrule=0.8pt, arc=3pt, boxsep=4pt,
    title={\textbf{Verification Walkthrough}}]
\scriptsize

\textbf{Sentence 1:} \textit{``We start with lo=0 and hi=3.''}\\
Extracted entities: \texttt{lo=0, hi=3}\\
Window scan (pointer at [1]): \texttt{lo=0} found at [3] \checkmark, \texttt{hi=3} found at [5] \checkmark. Pointer advances to [5].

\vspace{3pt}
\textbf{Sentence 2:} \textit{``mid = (0+3)//2 = 1. arr[1]=3 < 5, so lo becomes 2.''}\\
Extracted entities: \texttt{mid=1, arr[1]=3, lo=2}\\
Window scan: \texttt{mid=1} found at [8] \checkmark, \texttt{arr[1]<target} matches [10] \checkmark, \texttt{lo=2} found at [12] \checkmark. Pointer advances to [12].

\vspace{3pt}
\textbf{Sentence 3:} \textit{``mid = (2+3)//2 = 2. arr[2]=5.''}\\
Extracted entities: \texttt{mid=2, arr[2]=5}\\
Window scan: \texttt{mid=2} found at [15] \checkmark, \texttt{arr[2]==5} matches [16] \checkmark. Pointer advances to [16].

\vspace{3pt}
\textbf{Sentence 4:} \textit{``Since 5>5 is false and 5<5 is false, we enter the else branch and set hi=1.''}\\
Extracted entities: \texttt{\textcolor{red}{else branch}, \textcolor{red}{hi=1}}\\
Window scan ([16]--[18]+$K$): \textcolor{red}{Trace shows \texttt{if arr[2]==5 $\to$ True} at [16], followed by \texttt{return mid} at [17]. No \texttt{else branch} found. \texttt{hi=1} not found in window or accumulated state.} \textbf{Event Matching fails.}

\vspace{3pt}
\textbf{Sentence 5:} \textit{``Return -1.''}\\
Extracted entities: \texttt{\textcolor{red}{return=-1}}\\
\textcolor{red}{Trace shows \texttt{Return value: 2} at [18].} \textbf{I/O check also fails.}

\end{tcolorbox}

    \end{document}